\documentclass[aps, prd, reprint, superscriptaddress, nofootinbib, bibnotes]{revtex4-2}
\usepackage{graphicx}
\usepackage{dcolumn}
\usepackage{bm}
\usepackage{float}
\usepackage{lipsum}
\usepackage{xcolor}
\usepackage{textcomp}
\usepackage{latexsym}
\usepackage[colorlinks=true, allcolors=blue]{hyperref}
\usepackage{mathtools}
\usepackage{url}
\usepackage{soul}
\usepackage{comment}
\usepackage{xspace}
\usepackage{subfigure}
\usepackage{acronym}
\usepackage{resizegather}
\usepackage[T1]{fontenc}
\usepackage{ae,aecompl}
\usepackage[utf8]{inputenc}
\usepackage{graphicx, subfigure}        
\usepackage{amssymb,amsmath,placeins,epsfig,array,bm}
\usepackage{aas_macros}

\newcommand{\Mdot}{\mathrm{M}_\odot}

\newcommand{\red}[1]{{\color{black}{#1}}}

\begin{document}

\title{Cosmological feedback from a halo assembly perspective}

\author{Luisa Lucie-Smith}
\email[]{luisa.lucie-smith@uni-hamburg.de}
\affiliation{Hamburger Sternwarte, Universit{\"a}t Hamburg, Gojenbergsweg 112, D-21029 Hamburg, Germany}

\author{Hiranya V. Peiris}
\email[]{hiranya.peiris@ast.cam.ac.uk}
\affiliation{Institute of Astronomy and Kavli Institute for Cosmology, University of Cambridge, Madingley Road, Cambridge CB3 0HA, UK}
\affiliation{The Oskar Klein Centre for Cosmoparticle Physics, Department of Physics, Stockholm University, AlbaNova, Stockholm SE-106 91, Sweden}

\author{Andrew Pontzen}
\email[]{andrew.p.pontzen@durham.ac.uk}
\affiliation{Institute for Computational Cosmology, Department of Physics, Durham University, South Road, Durham, DH1 3LE, UK}

\author{Anik Halder}
\affiliation{Institute of Astronomy and Kavli Institute for Cosmology, University of Cambridge, Madingley Road, Cambridge CB3 0HA, UK}

\author{\\Joop Schaye}
\affiliation{Leiden Observatory, Leiden University, PO Box 9513, 2300 RA Leiden, the Netherlands}
\author{Matthieu Schaller}
\affiliation{Leiden Observatory, Leiden University, PO Box 9513, 2300 RA Leiden, the Netherlands}
\affiliation{Lorentz Institute for Theoretical Physics, Leiden University, PO Box 9506, NL-2300 RA Leiden, The Netherlands}
\author{John Helly}
\affiliation{Institute for Computational Cosmology, Department of Physics, Durham University, South Road, Durham, DH1 3LE, UK}
\author{Robert J. McGibbon}
\affiliation{Leiden Observatory, Leiden University, PO Box 9513, 2300 RA Leiden, the Netherlands}
\author{Willem Elbers}
\affiliation{Institute for Computational Cosmology, Department of Physics, Durham University, South Road, Durham, DH1 3LE, UK}

\newcommand{\AH}[1]{\textcolor{orange}{\textbf{#1}}}

\date{\today}
\begin{abstract}
The impact of feedback from galaxy formation on cosmological probes is typically quantified in terms of the suppression of the matter power spectrum in hydrodynamical compared to gravity-only simulations. In this paper, we instead study how baryonic feedback impacts halo assembly histories and thereby imprints on cosmological observables. We investigate the sensitivity of the thermal Sunyaev-Zel’dovich effect (tSZ) power spectrum, X-ray number counts, weak lensing and kinetic Sunyaev-Zel’dovich (kSZ) stacked profiles to halo populations as a function of mass and redshift. We then study the imprint of different feedback implementations in the FLAMINGO suite of cosmological simulations on the assembly histories of these halo populations, as a function of radial scale. We find that kSZ profiles target lower-mass halos ($M_{200\mathrm{m}}\sim 10^{13.1}~\mathrm{M}_\odot$) compared to all other probes considered ($M_{200\mathrm{m}}\sim 10^{15}~\mathrm{M}_\odot$). Feedback is inefficient in high-mass clusters with $\sim 10^{15} \, \mathrm{M}_\odot$ at $z=0$, but was more efficient at earlier times in the same population, with a  $\sim 5$--$10\%$ effect on mass at $2<z<4$ (depending on radial scale). Conversely, for lower-mass halos with $\sim10^{13}~\mathrm{M}_\odot$ at $z=0$, feedback exhibits a $\sim5$--$20\%$ effect on mass at $z=0$ but had little impact at earlier times ($z>2$). These findings are tied together by noting that, regardless of redshift, feedback most efficiently redistributes baryons when halos reach a mass of $M_{\rm 200m} \simeq {10^{12.8}}\,\mathrm{M}_{\odot}$ and ceases to have any significant effect by the time $M_{\rm 200m} \simeq {10^{15}}\,\mathrm{M}_{\odot}$. We put forward strategies for minimizing sensitivity of lensing analyses to baryonic feedback, and for exploring baryonic resolutions to the unexpectedly low tSZ power in cosmic microwave background observations. 
\end{abstract}

\maketitle

\section{Introduction}
Understanding the impact of baryonic feedback processes on cosmological observables remains one of the key challenges in modern cosmology \cite{Chisari_2019}. In particular, feedback due to active galactic nuclei (AGN) can redistribute matter within and beyond dark matter halos, leaving imprints on cosmological observables that are sensitive to the total matter and/or gas \cite{van_Daalen_2011, Eckert2021, Semboloni_2011}. Traditionally, this impact has been quantified through global summary statistics, such as the suppression of the matter power spectrum $P(k)$ in hydrodynamical simulations relative to gravity-only simulations \cite{Schaller2025}. However, such an approach may obscure the underlying physical mechanisms driving these effects. Physically, this suppression largely reflects the baryon mass fraction in halos as a function of their mass and radius \cite{Semboloni_2011, Semboloni_2013, Schneider_2019,vanDaalen2020,Debackere_2019,Arico2021,Salcido_2023,Grandis2024}.

In this work, we present a new perspective on baryonic feedback by examining it through the lens of halo mass assembly histories (MAH). Rather than focusing solely on the net effect on summary statistics, we investigate how feedback processes affect different halo populations over the course of their evolutionary histories. This approach allows us to establish a more direct connection between the cosmological observables and the imprint of feedback on halo populations at different masses, redshifts and halo radii.

Our analysis is particularly motivated by the apparent tension between baryonic feedback constraints derived from different cosmological probes.
Observationally, there appears to be a tension between the expected baryon fraction from X-ray measurements \cite{Sun_2009, Lovisari_2015, Akino_2022}, and that from detections of the kinetic Sunyaev-Zel'dovich (kSZ) effect around halos through stacking analyses \cite{Schaan2021, Hadzhiyska:2024qsl, Bigwood2024, mccarthy2024flamingocombiningkineticsz, Hadzhiyska2025PhRvD}. When compared to simulations, the latter have been interpreted to suggest a stronger impact of baryonic feedback compared to that predicted by simulations which have been calibrated to reproduce the gas fractions of low redshift X-ray-selected groups and clusters \cite{Hadzhiyska:2024qsl, Bigwood2024, mccarthy2024flamingocombiningkineticsz}.
One of the aims of this work is to show how these apparently contradictory results can be reconciled by examining the halo populations probed by each observable.

Previous work has connected baryonic feedback effects on cosmology with halo masses, for example by calculating the contribution of halos of a given mass range to the matter power spectrum \cite{van_Daalen_2015,vanLoon2024,Mead2020}, to the matter-electron pressure \cite{Mead2020} and to the cosmic shear signal in weak lensing \cite{To2024}. Similar studies were also done to decompose the thermal Sunyaev-Zel'dovich (tSZ) effect power spectrum \cite{Komatsu_2002, Makiya_2018, Battaglia2012, McCarthy2014} by halo mass and redshift. \citet{Elbers2024} also investigated the impact of baryonic feedback effects on halo mass accretion history by matching halos in the hydrodynamical and gravity-only versions of the same simulation setup. They studied how the ratio $M_{\rm hydro}/M_{\rm DMO}$ changes when selecting halos based on their concentration, formation time and large-scale environment. In the present work our focus is instead on understanding how the growth history of halos in hydrodynamical simulations determines the sensitivity of specific cosmological observables to baryonic feedback.

The paper is structured as follows. In Sec.~\ref{sec:background}, we set the context on observational probes of baryonic feedback, and in Sec.~\ref{sec:sims} we describe the simulations used in this work. In Sec.~\ref{sec:probes} we present our methodology for determining the halo mass and redshift ranges probed by various cosmological observables, namely the tSZ effect, weak lensing, X-ray, and the kSZ effect.
Our results are then presented in Sec.\ref{sec:results}. We begin in Sec.\ref{sec:probes_sensitivity} with a comparison of how the different probes respond to the imprint of feedback on halos across mass and redshift.
In Sec.~\ref{sec:feedback} we then show the imprint of feedback on the evolution histories of halo populations targeted by each observable. We further investigate the impact of feedback in terms of fixed time intervals in Sec.\ref{sec:feedbackfixedtime}, and its radial dependence in Sec.~\ref{sec:feedbackradial}. 
We discuss the implications of our work and conclude in Sec.~\ref{sec:conclusions}.

\section{Background}
\label{sec:background}
X-ray observations have traditionally provided the highest quality measurements of the properties of hot gas in groups and clusters \cite{Bulbul2024, Popesso2024}. When complemented by total mass measurements -- either from X-ray observations themselves, dynamics or weak lensing -- they can be used to infer the hot gas mass fraction \cite{Sun_2009, Lovisari_2015, Akino_2022}. Since this hot gas is redistributed by energetic feedback processes, X-ray measurements offer a powerful way to constrain the efficiency of baryonic feedback, particularly in the inner regions of groups and clusters. Therefore, the inferred gas mass fraction from X-ray measurements is employed in hydrodynamical simulations such as BAHAMAS \cite{McCarthy_2016} and FLAMINGO \cite{Schaye_2023} to help calibrate the sub-grid models controlling the efficiencies of feedback in simulations.

The tSZ and kSZ effects around groups and clusters also yield valuable information on the gas around halos. The tSZ effect arises from the inverse Compton scattering of  cosmic microwave background (CMB) photons by the hot free electrons in groups and clusters. The amplitude of the effect -- typically expressed in terms of the Compton-$y$ parameter -- is directly proportional to the electron pressure integrated along the line-of-sight \cite{Sun_2009}. The tSZ effect therefore traces hot ionized gas in the Universe, which is in turn correlated with the large-scale structure. Thus, the tSZ effect contains useful cosmological information \cite{Planck2016tSZ, Bolliet2018, Makiya_2018}, in particular for constraining the amplitude of density perturbations $\sigma_8$ \cite{Komatsu_2002}, and provides an important constraint on the magnitude of baryonic feedback due to its dependence on the gas content \cite{Schneider_2022, McCarthy2014}. 

Recent measurements of the tSZ power spectrum have been made by \textit{Planck} \cite{Planck2016tSZ, Bolliet2018} on large angular scales, and by the Atacama Cosmology Telescope (ACT; \cite{Choi2020}) and the South Pole Telescope (SPT; \cite{Reichardt2021}) on smaller angular scales. Both of these have been reported to be systematically lower than predictions based on $\Lambda$CDM.
The discrepancy between the measurements and the predictions of the tSZ power spectrum from simulations and its cross spectrum with cosmic shear cannot be resolved through increased feedback alone. On large scales, the discrepancy appears to be more easily alleviated with changes to the cosmology rather than the strength of feedback, while on small scales feedback can have a significant effect on the shape of the tSZ power spectrum \cite{McCarthy2014, McCarthy_2018, mccarthy2024flamingocombiningkineticsz, efstathiou2025}.

The kSZ effect arises instead from the interaction between CMB photons and free electrons in bulk motion relative to the CMB rest-frame. This effect depends on the integrated electron density along the line-of-sight when combined with estimates of the peculiar velocity, thus offering a direct probe of the spatial distribution and abundance of baryons even out to the outskirts of galaxies and clusters. 

Detections of the kSZ effect around halos through stacking analyses are relatively new \cite{PhysRevD.93.082002, Hand_2012, 2016A&A...586A.140P, PhysRevLett.117.051301, Amodeo_2021}. Recent kSZ measurements, combined with CMB data from ACT with individual velocity estimates from the CMASS catalog of the Baryon Oscillation Spectroscopic Survey (BOSS; \cite{Ahn2014}), showed that the gas density profiles in groups deviate significantly from the Navarro-Frenk-White (NFW \cite{Navarro1997}) radial profile expected in the absence of feedback. More recent studies of a \textit{Planck}+ACT kSZ effect stacking analysis of galaxies in the Dark Energy Spectroscopic Instrument (DESI) Legacy Imaging Survey \cite{Hadzhiyska:2024qsl} confirmed the finding of much more extended gas profiles than expected from the dark matter radial profiles. When compared to simulations, the authors assert that strong feedback is required to match their stacked profiles and that the original Illustris simulation (which indeed adopts a strong feedback model) provides a good fit to the data, while the more recent Illustris-TNG feedback model is ruled out. 

Recent efforts to constrain both baryonic feedback and cosmology by jointly modeling the \citet{Schaan2021} kSZ effect and cosmic shear correlation functions using baryonification models have pointed to a similar picture \cite{Bigwood2024}: the data prefer a stronger impact of baryons than predicted by simulations which have been calibrated to reproduce the gas fractions of low redshift X-ray-selected groups and clusters. This was further confirmed by \citet{mccarthy2024flamingocombiningkineticsz} who compared the recent kSZ measurements to predictions from the FLAMINGO hydrodynamical simulations, finding that more aggressive feedback is required in the simulations in order to fit the data compared to that inferred using X-ray cluster observations.

\section{Simulations}
\label{sec:sims}
We use the FLAMINGO \cite{Schaye_2023} suite of simulations to study the connection between the evolutionary history of halos and baryonic feedback. The simulations were run with the cosmological smoothed particle hydrodynamics and gravity code SWIFT \cite{Schaller_2024} using the SPHENIX SPH scheme \cite{Borrow2022}, starting from initial conditions generated with a modified version of \texttt{monofonIC} \cite{Hahn:2020lvr, Elbers_2022}. The suite
has variations in resolution, box size, cosmology, and subgrid modeling. 

In this work, we consider variations to the subgrid modeling, while keeping fixed the cosmology, resolution and box size. We start with the gravity-only and hydrodynamical fiducial runs of box size $L=1 \, \mathrm{Gpc}$ and resolution $N=1800^3$ (the resolution is equivalent to both the number of baryonic and cold dark matter particles). In the hydrodynamical run, the baryonic and cold dark matter particle masses are $m_{\rm b} = 1.07 \times 10^9 \, \Mdot$ and $m_{\rm cdm} = 5.65 \times 10^9 \, \Mdot$; in the gravity-only run, the cold dark matter particle mass is $m_{\rm cdm} = 6.72 \times 10^9 \, \Mdot$. The comoving Plummer-equivalent gravitational softening is $\epsilon_{\rm com}=22.3 \, \mathrm{kpc}$, and the maximum proper gravitational softening is $\epsilon_{\rm prop}=5.7 \, \mathrm{kpc}$. The cosmological parameters are those corresponding to the DES Y3 `3x2pt + All Ext.' $\Lambda$CDM cosmology \cite{2022PhRvD.105b3520A}, which assume a spatially flat universe and are based on the combination of constraints from DES Y3 `3$\times$2 point correlation function and from external data from BAO, redshift-space distortions, SN Type Ia, and \textit{Planck} observations of the CMB (including CMB lensing), Big Bang nucleosynthesis, as well as local measurements of the Hubble constant.

In addition to the fiducial simulations, we consider two simulations which implement enhanced feedback. In FLAMINGO, there are two types of observational data used to calibrate the subgrid feedback models which contain free parameters: the $z=0$ stellar mass function (SMF) and the gas fractions in groups and clusters, $f_{\rm gas}(M_{\rm 500c})$\cite{Kugel2023}. The subgrid model of the fiducial simulations is calibrated to reproduce a compilation of observations of these quantities. 
In particular, the gas fraction measurements come from the compilation of $z\approx0.1$ X-ray data from \citet{Kugel2023} and the $z\approx0.3$ weak lensing (plus X-ray) data from \citet{Akino_2022,Mulroy2019,Hoekstra2015}.

To construct simulations with varying feedback scenarios, the subgrid free parameters are recalibrated to find models in which the cluster gas fractions and/or the SMF have higher/lower values than the observationally preferred range. In this work, we use the `fgas$-8\sigma$' simulation, where  the gas fraction as a function of halo mass in groups and clusters is shifted lower by $8\sigma$; the shifted data is then used to recalibrate a new set of subgrid parameters. 
We additionally consider a FLAMINGO variation  which changes the AGN feedback implementation from thermal to an anisotropic, kinetic jet-like feedback, which is then calibrated to a `$f_{\rm gas}-4\sigma$' relation. This simulation is denoted `Jet\_fgas$-4\sigma$'.

\subsection{Halo identification and properties}
Cosmic structures were identified using HBT-HERONS \cite{ForouharMoreno2025} a recently updated version of the Hierarchical Bound Tracing algorithm (HBT+; \cite{Han:2017lpe}). This algorithm identifies structures as they form across time, so that at each step particles are grouped via a friends-of-friends (FOF) algorithm combined with an iterative unbinding procedure. The particles associated to these self-bound objects are tracked across simulation outputs, thus yielding a set of candidate substructures at subsequent times. This additionally allows the identification of satellites within more massive halos. We further make use of the Spherical Overdensity and Aperture Processor (SOAP; \cite{McGibbon_2025SOAP}), which computes a number of halo properties in a range of apertures. We will make use of halo properties computed via SOAP throughout this work.

In all simulations we consider, we use three different mass definitions: $M_{\rm 500c}$, i.e., the mass enclosed within a sphere of density $500\times \rho_{\rm crit}$ where $\rho_{\rm crit}$ is the critical density of the Universe, $M_{\rm 200m}$, i.e., the mass enclosed within a sphere of density $200\times \rho_{\rm m}$ where $\rho_{\rm m}$ is the mean matter density in the Universe, and $M_{\rm 5\times500c}$, i.e., the mass enclosed within a sphere of radius $r=5 \times r_{\rm 500c}$. In all simulations, we consider primarily halos with total mass $M_{\rm 200m} \geq 10^{13}~\Mdot$. 

We further connect the halos across the gravity-only and hydrodynamical counterpart simulations by matching the ten most strongly-bound particles of halos between the two simulations. This allows us to compare the properties of the same halos (including their accretion histories) with and without baryonic effects.

As the HBT-HERONS algorithm finds substructures by tracking them across simulation outputs, it naturally also provides a self-consistent way to define the mass assembly histories (MAH) of the halos. The MAH is given by the mass of the halo as it is tracked (and identified) across all simulation snapshots. 

\begin{figure*}
    \centering
    \includegraphics[width=0.95\textwidth]{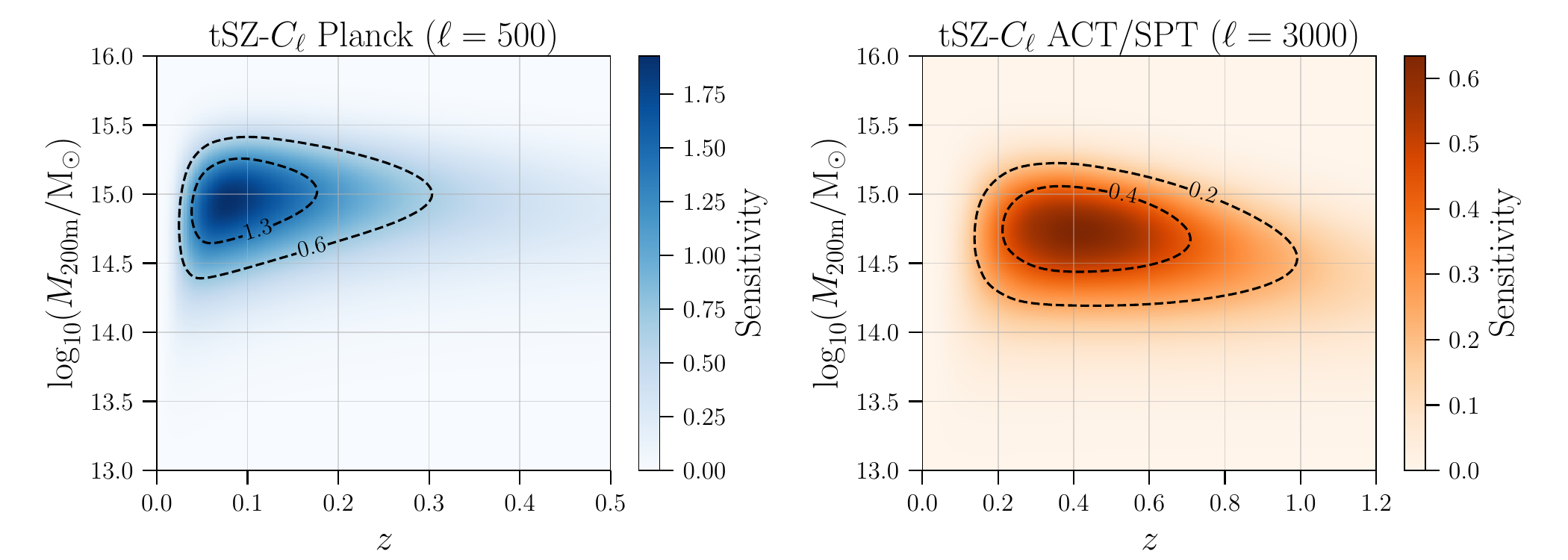}
    \caption{The sensitivity of the tSZ power spectrum to halos as a function of their mass and redshift, as defined by Eq.~\eqref{eq:tSZ-sensitivity}. The left panel shows the sensitivity at $\ell=500$, the mid-point of the $\ell$-range probed by \textit{Planck} \cite{Planck2016tSZ, Bolliet2018}; the right panel shows that for $\ell=3000$, indicative of the scales probed by ACTpol \cite{Choi2020} and SPT \cite{Reichardt2021}. The inner and outer dashed contours are isocontours which correspond to respectively 33\% and 66\% of the maximum sensitivity of the signal.}
    \label{fig:tsz_sensitivity}
\end{figure*}

\section{Cosmological probes}
\label{sec:probes}
We now consider different cosmological probes -- tSZ, weak lensing, X-ray, kSZ -- and describe the methodology used to compute their sensitivity to halos as a function of their mass and redshift.

\subsection{Mass definition}
\label{sec:massdef}

Different cosmological probes are sensitive to different radial scales within the halo. This leads to observational measurements being reported in terms of different mass definitions tied to different halo radii. X-ray observations primarily probe the inner regions of clusters, and therefore report estimated cluster masses in terms of $M_{\rm 500c}$.  Cosmic shear, on the other hand, probes the integrated mass of halos over a wide range of masses and redshift, and is therefore sensitive to scales enclosing $M_{\rm 200m}$. To compare measurements from different cosmological probes on a similar footing, we convert between different spherical overdensity mass definitions assuming a fixed functional form for the density profile as a function of radius in physical units at any given redshift. This is given by the Navarro-Frenk-White (NFW) density profile \cite{NFW1997}, where its free parameters are fixed by the concentration-to-mass relation model $c(M)$ of \citet{DiemerJoyce2019}. The latter improves the original $c(M)$ relation of \citet{Diemer2015} and has been tested on different mass definitions and different redshifts. We tested our analytic model for converting between mass definitions against the FLAMINGO simulation data and found it provides a good fit, based on comparisons of halo mass definitions at several selected redshifts.

\subsection{Thermal Sunyaev-Zel'dovich sensitivity analysis}
Cosmology from the tSZ effect is studied through the angular power spectrum $C_\ell^{\rm tSZ}$ of the Compton-$y$ field, which is widely formulated within the halo-model framework \cite{Refregier_2002}. For multipoles $\ell > 100$, the one-halo term provides the dominant contribution to $C_\ell^{\rm tSZ}$ \cite{Komatsu_1999, Komatsu_2002}. Following \citet{Bolliet2018} and assuming a halo model, the tSZ power spectrum (one-halo term) can be written as:
\begin{equation}
    C_\ell^{\rm tSZ} = \int \mathrm{d}z \frac{\mathrm{d}V}{\mathrm{d}z\,\mathrm{d}\Omega} \int \mathrm{d}M \frac{\mathrm{d}n(M, z)}{\mathrm{d}M} \, | \tilde{y}_{\ell}(M,z) |^2 \ ,
\end{equation}
where $V(z)$ is the comoving volume of the universe, $\mathrm{d}n / \mathrm{d}M$ is the halo mass function, and $y_{\ell}$ the two-dimensional Fourier transform of the line-of-sight projected Compton $y$-parameter. We use \texttt{class\_sz} \cite{bolliet2023classszioverview} to estimate the sensitivity of the tSZ power spectrum $\mathcal{C}_\ell$ to halos of different masses and redshifts. We define the sensitivity (or response) of $C_\ell^{\rm tSZ}$ to redshift and mass for a given $\ell$ through the second derivative of $C_\ell^{\rm tSZ}$ as follows:\footnote{We suppress the explicit dependence of the functions on $(M,z)$ for the sake of succinct notation.}
\begin{align}
    \frac{1}{C_\ell^{\rm tSZ}} \frac{\mathrm{d}^2{C_\ell^{\rm tSZ}}}{\mathrm{d}z \, \mathrm{d}\ln M} &= \frac{M}{C_\ell^{\rm tSZ}} \frac{\mathrm{d}^2}{\mathrm{d}z \, \mathrm{d}M} \int \mathrm{d}z \frac{\mathrm{d}V}{\mathrm{d}z\,\mathrm{d}\Omega} \int \mathrm{d}M \frac{\mathrm{d}n}{\mathrm{d}M} \,  | \tilde{y}_{\ell} |^2 \ \\
    &= \frac{M}{C_\ell^{\rm tSZ}} \frac{\mathrm{d}V}{\mathrm{d}z\,\mathrm{d}\Omega} \frac{\mathrm{d}n}{\mathrm{d}M} \,  | \tilde{y}_{\ell}|^2,\label{eq:tSZ-sensitivity}
\end{align}
similar to that presented in \citet{Komatsu_2002}. We perform the calculation in terms of $M_{\rm 500c}$ and then convert it to $M_{\rm 200m}$ according to the procedure described in \S~\ref{sec:massdef}. We consider a Tinker halo mass function \cite{Tinker_2008} and a generalized NFW pressure profile \cite{Nagai2007}. The \textit{Planck} analysis reports measurements of the tSZ power spectrum in the range $\ell \in [100, 1000]$ \cite{Planck2016tSZ, Bolliet2018}, while measurements from the SPT data from \citet{Reichardt2021} and the ACTpol measurement from \cite{Choi2020} are reported at $\ell \sim 3000$, as ACT and SPT have almost an order of magnitude better angular resolution and higher sensitivity
than \textit{Planck}. 
We therefore consider the sensitivity of \textit{Planck} using a representative scale $\ell=500$, while for ACT and SPT we compute the sensitivity at $\ell=3000$. 

We illustrate the tSZ-sensitivity in Fig.~\ref{fig:tsz_sensitivity} for \textit{Planck} (left panel), and ACT and SPT (right panel); the inner and outer dashed contours have respectively 33\% and 66\% of the maximum sensitivity of the signal. \textit{Planck} probes high-mass, rare clusters of $M_{\rm 200m}\sim 10^{14.6-15.4}~\Mdot$ at $z<0.3$, while the sensitivity of ACT/SPT is dominated by $M_{\rm 200m}\sim 10^{14.4-15.1}~\Mdot$ halos across a higher redshift range $z\sim 0.2 - 1$. Our results are qualitatively consistent with previous work which explored mass and redshift as independent variables \cite{Komatsu_2002, Makiya_2018}, but show that the redshift and mass sensitivities are not fully separable as previously pointed out in \citet{Holder2007, Battaglia2012}. 

\subsection{Weak lensing sensitivity analysis}
\begin{figure*}
    \centering
    \includegraphics[width=0.9\textwidth]{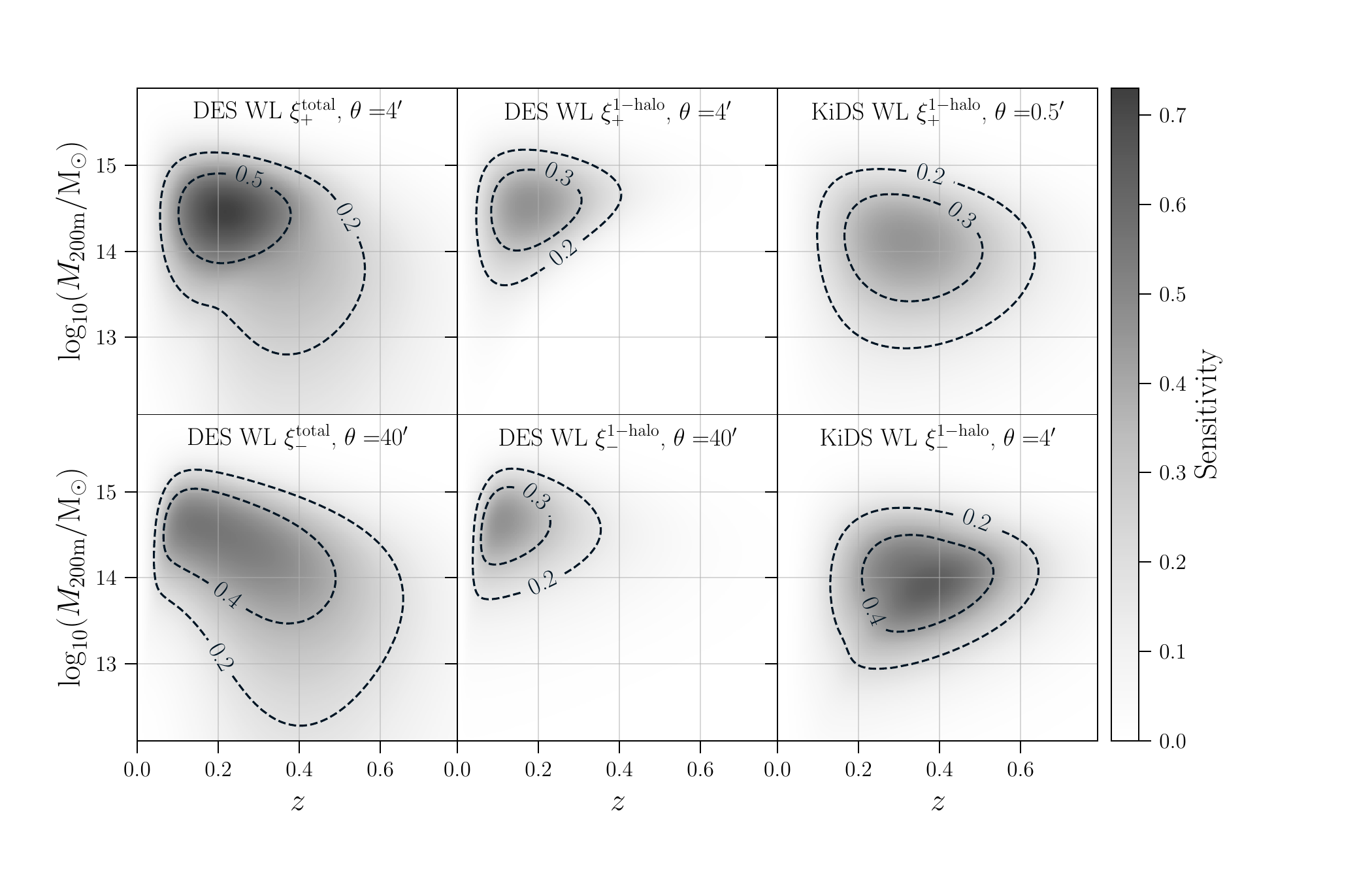}
    \caption{The sensitivity of cosmic shear two point correlation function, $\xi_{+}$ (top row) and $\xi_{-}$ (bottom row), as defined by Eq.~\eqref{eq:lensing-sensitivity-1h2h} (left column) and Eq.~\eqref{eq:lensing-sensitivity-1h-only} (middle and right columns). From left to right the columns show: the total sensitivity of the smallest angular scales in DES \cite{Secco_2022, Amon_2022} analyses ($\theta=4'$ for $\xi_+$ and $\theta=40'$ for $\xi_-$); the one-halo sensitivity of the same angular scales (i.e., the same scales but now showing only the term which can change due to gas redistribution); and the one-halo sensitivity of the smallest angular scales used in the KiDS \cite{Asgari_2021} analyses ($\theta=0.5'$ for $\xi_{+}$ and $\theta=4'$ for $\xi_{-}$). We adopt the surveys' highest redshift source tomographic bins, namely bin 4 for DESY3 and bin 5 for KiDS-1000.} 
    \label{fig:WL_sensitivity}
\end{figure*}
Constraints on cosmological parameters from galaxy weak lensing are obtained by analyzing the two-point correlation functions $\xi_+(\theta)$ and $\xi_-(\theta)$. These correspond to correlations of tangential and cross components of the shear field relative to the line separating two galaxies at an angular distance of $\theta$. The $\xi_{\pm}(\theta)$ can be written as a line-of-sight projection of the 3D matter power spectrum $P_{\delta}(k,z)$ with the corresponding projection kernel $q(\chi)$, the lensing efficiency for a single source galaxy tomographic bin:
\begin{align}
    \xi_{\pm}(\theta) &= \int \frac{\rm d \ell \, \ell}{2\pi} \, J_{0/4}(\ell \theta) \int \frac{c\ \mathrm{d} z}{H(z)} \frac{q(\chi)^2}{\chi^2} P_{\delta} \left(k=\frac{\ell}{\chi}, z \right)\,,
\end{align}
where $J_{0/4}$ are the 0th and 4th order ordinary Bessel functions of the first kind, and, $\chi(z), H(z)$ are the comoving distance and the Hubble parameter evaluated at redshift $z$, respectively. Within the context of the halo model, the $\xi_{\pm}$ can be described by a combination of one-halo $P_{\rm 1h}$ and two-halo $P_{\rm 2h}$ contributions to the matter power spectrum $P_{\delta} = P_{\rm 1h} + P_{\rm 2h}$ \cite{COORAY_2002, Asgari_2023}, where:
\begin{align}
\label{eq:1_halo_term}
    P_{\text{1h}}(k,z) &= \frac{1}{\bar{\rho}^2}\int \mathrm{d}M \, M^2 \frac{\mathrm{d}n}{\mathrm{d}M} \, \tilde{u}(k|M,c)^2 \textrm{ and} \\
    P_{\text{2h}}(k,z) &= P_{\text{lin}}(k,z) \nonumber \\
     & \quad \times \left[ \frac{1}{\bar{\rho}}\int \mathrm{d}M \, M \frac{\mathrm{d}n}{\mathrm{d}M} \, b(M,z) \tilde{u}(k|M,c) \right]^2 \ .
\end{align}
In the equations above the various symbols stand for the following: $\mathrm{d}n / \mathrm{d}M$ is the halo mass function, $\tilde{u}(k|M;c)$ is the Fourier transform of the normalized halo density profile, $c$ the concentration, $\bar{\rho}$ the mean matter density of the universe, $b(M,z)$ the linear halo bias, and $P_{\text{lin}}(k,z)$ the linear matter power spectrum. We enforce mass conservation relations such that the large-scale halo-model matter power spectrum obeys the standard linear perturbation theory prediction. Combining these equations, we can compute the sensitivity (second derivative) of $\xi_{\pm}$ for a given angular bin $\theta$ with respect to mass and redshift as:
\begin{equation}
\begin{split}
    \frac{1}{\xi_{\pm}(\theta)}\frac{\mathrm{d}^2\xi_{\pm}(\theta)}{\mathrm{d}z \, \mathrm{d}\ln M} &= \frac{M}{\xi_{\pm}(\theta)} \frac{\mathrm{d}^2}{\mathrm{d}z \, \mathrm{d}M} \Bigg[ \int \frac{\rm d \ell \, \ell}{2\pi} \, J_{0/4}(\ell \theta) \int \frac{c\ \mathrm{d} z}{H(z)} \frac{q(\chi)^2}{\chi^2} \\
    &\qquad \qquad \qquad \times \Big( P_{\text{1h}} (\ell/\chi, z) + P_{\text{2h}} (\ell/\chi, z) \Big) \Bigg] \\
    &= \frac{M}{\xi_{\pm}(\theta)}  \int \frac{\rm d \ell \, \ell}{2\pi} \, J_{0/4}(\ell \theta) \frac{c}{H(z)} \frac{q(\chi)^2}{\chi^2} \Bigg[\frac{\mathrm{d} P_{\text{1h}}}{\mathrm{d} M} + \frac{\mathrm{d} P_{\text{2h}}}{\mathrm{d} M} \Bigg] \ ,\label{eq:lensing-sensitivity-1h2h}
\end{split}
\end{equation}
with the mass derivatives of the one-halo and two-halo power spectra in turn given by :
\begin{equation}
\begin{split}
    \frac{\mathrm{d} P_{\text{1h}}(k,z)}{\mathrm{d} M} = \frac{1}{\bar{\rho}^2} \ M^2 \frac{\mathrm{d}n}{\mathrm{d}M} \tilde{u}(k|M,c)^2 \ ,
\end{split}
\end{equation}
and
\begin{equation}
\begin{split}
    \frac{\mathrm{d} P_{\text{2h}}(k,z)}{\mathrm{d} M} & = P_{\text{lin}}(k,z) \frac{2}{\bar{\rho}}M  \frac{\mathrm{d}n}{\mathrm{d}M} b(M,z) \tilde{u}(k|M,c) \\
    & \qquad \times \left[ \frac{1}{\bar{\rho}}\int \mathrm{d}M' \ M'  \frac{\mathrm{d}n}{\mathrm{d}M} b(M',z) \tilde{u}(k|M',c) \right] \ ,
\end{split}
\end{equation}
where we denote $k = \ell/\chi$. As for the tSZ case, we consider a Tinker halo mass function \cite{Tinker_2008}, a Tinker halo bias \cite{Tinker_2010}, and the concentration-to-mass relation model $c(M)$ of \citet{DiemerJoyce2019}.

In Fig.~\ref{fig:WL_sensitivity} we show the sensitivity of $\xi_{\pm}$ for the total and the one-halo contributions at the smallest angular scales used by two Stage-III weak lensing surveys for their highest redshift source tomographic bins,\footnote{We consider the highest redshift tomographic bins as they carry the highest signal-to-noise of the cosmic shear signal.} namely, bin 4 of the DESY3 (Year 3 cosmic shear analysis of the Dark Energy Survey \cite{Amon_2022, Secco_2022}), and bin 5 of KiDS-1000 (the cosmic shear analysis of the fourth data release of the Kilo Degree Survey \cite{Asgari_2021}). The upper and lower panels show $\xi_{+}$ and $\xi_{-}$ respectively.
The smallest scales (after application of scale-cuts) used in the DESY3 analysis  are $\theta \sim 4'$ for $\xi_{+}$ and $\theta \sim 40'$ for $\xi_{-}$; for tomographic bin 4 of DESY3. By contrast, the smallest angular scales used in the KiDS-1000 analysis are $\theta\sim0.5'$ for $\xi_{+}$ and $\theta\sim4'$ for $\xi_{-}$ for tomographic bin 5 of KiDS-1000. 

Feedback mainly redistributes matter on small scales within the one-halo regime while the two-halo contribution to $\xi_{\pm}$ probes the matter correlation function mainly on larger angular scales where the signal is governed by large-scale gravitational collapse (and the imprint of baryonic feedback is expected to be less significant).
While a wide range of halo masses contributes to the total lensing signal, on the small angular scales of $\xi_{\pm}(\theta)$ the signal is dominated by the one-halo term, probing a narrower range of halo masses. Hence, in Fig.~\ref{fig:WL_sensitivity} we show the relative sensitivity of the total two-point correlation function (left column) and one-halo term alone (middle column) for DES, and the sensitivity of the one-halo term only for the KiDS smallest scales (right column). Here, the contribution from the one-halo term is computed as a fraction of the total (one + two-halo) $\xi_{\pm}$ signal:
\begin{equation}
\begin{split}
    \frac{1}{\xi_{\pm}(\theta)}\frac{\mathrm{d}^2\xi^{\rm 1-halo}_{\pm}(\theta)}{\mathrm{d}z \, \mathrm{d}\ln M} &= \frac{M}{\xi_{\pm}(\theta)} \int \frac{\rm d \ell \, \ell}{2\pi} \, J_{0/4}(\ell \theta) \frac{c}{H(z)} \frac{q(\chi)^2}{\chi^2} \Bigg[\frac{\mathrm{d} P_{\text{1h}}}{\mathrm{d} M}  \Bigg]\ .
    \label{eq:lensing-sensitivity-1h-only}
\end{split}
\end{equation}
For completeness, in Fig.~\ref{fig:WL_sensitivity_appendix} of Appendix \ref{app:WL}, we show the $\xi_{\pm}$ sensitivity at a range of angular scales, and the corresponding contributions from the total, two-halo and one-halo terms for both DESY3 and KiDS-1000 highest-redshift source tomographic bins.\\
\\
We find that the one-halo term is sensitive to halos of $M_{\rm 200m}\sim 10^{14-15}~\Mdot$ at $z<0.4$ for DES, whereas at the smallest scales used in KiDS, the sensitivity shifts to a lower halo mass range at higher redshifts ($M_{\rm 200m}\sim 10^{13.5-14.5}~\Mdot$ at $z\sim0.2-0.6$). The sensitivity of the total $\xi_{\pm}$ extends to lower-mass halos than the one-halo contribution; this is because low-mass halos are far more abundant than higher mass halos and therefore dominate the two-halo term (see Appendix \ref{app:WL}). 

Our analysis is consistent with \citet{To_2024} (see their Fig. 3) which presents the accumulated contribution to the total $\xi_{\pm}$ signal assuming different mass-cuts to the one-halo term. For the smallest scales used in the DESY3 analysis, they find that halos with masses\footnote{Note that \citet{To_2024} report their halo masses in $\Mdot / h$ which we convert to $\Mdot$ assuming $h = 0.67$.} above $M_{\rm 200m}\sim 10^{13.7}~\Mdot$ carry the bulk of the one-halo signal in both $\xi_{\pm}$. This is consistent with our findings (see the lower extent of the 66\% contours in the one-halo term in Fig.~\ref{fig:WL_sensitivity}).

\subsection{eROSITA X-ray cluster catalog}
We consider the `cosmology sample' of the Cluster Catalog from the first Western All-Sky Survey of eROSITA (eRASS1) \cite{Bulbul2024, Ghirardini2024}. This is a purer sample of clusters compared to the entire eRASS1 catalog, which is more suitable for accurate determination of the cosmological parameters and testing of cosmological models. The sample is selected by applying cuts to the original Cluster Catalog in terms of the X-ray extent likelihood selection, $\mathcal{L}_\mathrm{ext} > 6$, the sky region, Dec $\leq 32.5$, and the redshift, $0.1<z<0.8$. The catalog is publicly available~\footnote{\href{https://erosita.mpe.mpg.de/dr1/AllSkySurveyData_dr1/Catalogues_dr1/}{
erosita.mpe.mpg.de/dr1/AllSkySurveyData\_dr1/Catalogues\_dr1}} and reports the redshifts and the estimated $M_{\rm 500c}$ values of the clusters. The mass was estimated using the scaling relations between count-rate and measured weak lensing shear profiles. We refer the reader to \citet{Bulbul2024} for a discussion on possible systematic effects and contamination in X-ray detected sources.

We compute the number density of halos in mass and redshift bins, smooth this using a Gaussian Mixture Model (GMM) from the publicly available code \texttt{GMM-MI} \cite{Piras_2023}, and report the 68\% and 95\% confidence region in Fig.~\ref{fig:M200m_sensitivity}.

\subsection{Stacked kinetic Sunyaev-Zel'dovich profiles}\label{sec:ksz}
Stacked analyses of the kSZ effect rely on the combination of high-resolution CMB maps (for example from -- \textit{Planck}, ACT and SPT -- with galaxy catalogs from spectroscopic (or photometric) redshift surveys such as BOSS or DESI. This allows one to stack the CMB temperature maps at the positions of the galaxies, weighting the stacks by the line-of-sight velocities of the galaxies.
To compute the sensitivity of the kSZ measurement, we consider as a representative example the galaxy sample used in \citet{Hadzhiyska:2024qsl} for the kSZ measurements i.e., the DESI `Main LRG' and `Extended LRG' sample with photometric redshifts described in \citet{Zhou_2023,Zhou_2023b}. These sources were selected from the imaging data from the DESI Legacy Imaging Survey Data Release 9 \cite{2019AJ....157..168D, Zhou_2023b}.

We adopt a similar selection cut to the one used in \citet{Hadzhiyska:2024qsl} (in turn following \citet{Zhou_2023}), including selecting galaxies with photo-$z$ in the range 0.4 < $z$ < 1.024 and with an estimated error of $\sigma(z) < 0.05$. This removes a small fraction of galaxies with anomalously large redshift errors. From this selection, we find that the galaxies in the DESI LRG sample probe a narrow range in stellar mass $\log (M_\star/\Mdot) \sim 11 - 11.5$ spread across redshifts $z\sim 0.4 - 0.9$; see Appendix~\ref{app:SMHM}. There exists many alternative methods for converting stellar masses to halo masses using galaxy-halo connection models \cite{WechslerTinker2018} such as abundance matching (e.g., \cite{Behroozi2010, Moster2010}), or weak lensing \cite{mccarthy2024flamingocombiningkineticsz}.

In order to link the stellar mass of a galaxy $M_\star$ at a given redshift $z$ to the mass of its dark matter halo $M_{\rm 200m}$, we use the empirical stellar-mass-to-halo-mass relation from the fiducial FLAMINGO simulation. We consider the stellar mass of central galaxies and their corresponding halo masses $M_{\rm 200m}$ at a representative redshift of $z=0.7$ in the FLAMINGO fiducial simulation. For every stellar mass value of the DESI LRG sample, we select a narrow bin around that value in the simulation and sample from the distribution of $M_{\rm 200m}$ halos within that stellar mass bin. This process is repeated for all stellar mass values of the DESI LRG sample. This yields a corresponding $M_{\rm 200m}$ value for each galaxy, which reflects the stellar-to-halo mass relation for central galaxies in FLAMINGO including the scatter in halo mass at fixed stellar mass. 

A small fraction of these galaxies ($11\% \pm 1\%$) \cite{Yuan2024} may in fact be satellites, and this has been shown to have no significant impact on the kSZ analysis by \citet{Hadzhiyska:2024qsl} using this catalog. To ensure that this also does not have a qualitative impact on our conclusions, we verify that including satellites from FLAMINGO and assigning the halo mass of the central has little overall impact, yielding only a small upward shift (0.1 -- 0.2 dex) in the inferred halo mass distribution of the DESI LRGs. We also test that our results do not change if we change the representative redshift to a different value within the redshift range of the DESI LRGs. This is expected since the stellar-to-halo mass relation within this redshift range is roughly redshift-independent.
We provide further details on the stellar to halo mass conversion in Appendix~\ref{app:SMHM}.

The sensitivity of the kSZ-galaxy cross-correlation measurements to halos in terms of their mass and redshift is proportional to the kSZ profile weighted by the number density of halos. 
We approximate the sensitivity of the kSZ signal as being given by the number density of the DESI LRG sample as a function of their halo mass and redshift, weighted by the gas mass of the halos within a fixed aperture physical scale of 3 Mpc at $z=0.7$. This closely matches the aperture scale used to measure the stacked kSZ profiles in \citet{Hadzhiyska:2024qsl}. We find no significant difference when adopting smaller fixed scale apertures even down to 50 kpc. This gas mass weighting (at fixed aperture) is a good proxy for the dependency of the kSZ signal on the gas content\footnote{Another weighting choice advocated in the literature is the spherical overdensity halo mass; however, this incorrectly introduces a dependence on the virial radius, to which the observational stacking procedure is not sensitive.}.

\begin{figure*}
    \centering
    \includegraphics[width=0.9\textwidth]{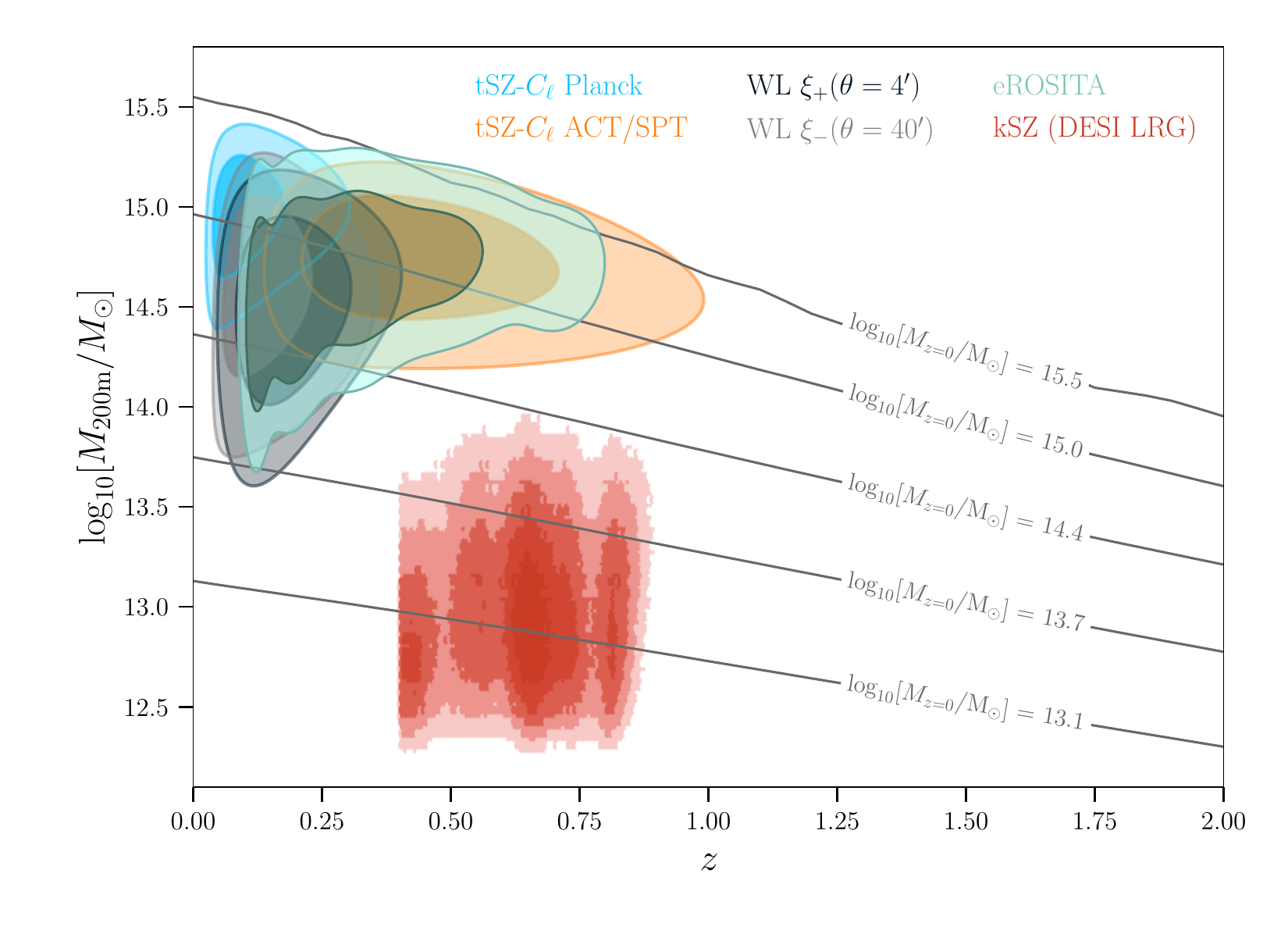}
    \caption{Sensitivity of different cosmological probes to the imprint of feedback on halo populations. As illustrative probes we choose the tSZ power spectrum at $\ell=500$ (\textit{Planck} \cite{Planck2016tSZ, Bolliet2018}) and $\ell=3000$ (ACT \cite{Choi2020} and SPT \cite{Reichardt2021}), as well as the one-halo contribution to the weak lensing two-point correlation function $\xi_{\pm}$ (for DESY3 source tomographic bin 4 \cite{Secco_2022, Amon_2022}), to halos as a function of their mass and redshift. In addition, we show halo number counts from the eROSITA X-ray catalog \cite{Bulbul2024} and from the photometric DESI LRG from the DESI Legacy
Imaging Surveys \cite{2019AJ....157..168D, Zhou_2023b, Zhou_2023}, which has been used to produce stacked kSZ profiles. Overplotted as grey lines are the median mass growth histories for halos of specified $z=0$ mass ($M_{z=0}$) in the fiducial FLAMINGO simulation, to aid understanding the relationship between different probes.}
    \label{fig:M200m_sensitivity}
\end{figure*}

Due to the complex dependency of the DESI LRG sample on the redshift, we choose to not smooth the number density distribution (e.g., with a GMM fit) to avoid misrepresenting the distribution. Instead, we show the data distribution directly using a 2D hexagonal binning plot in Fig.~\ref{fig:M200m_sensitivity}.

\section{Results}
\label{sec:results}

\subsection{Sensitivity of cosmological probes to halo populations}
\label{sec:probes_sensitivity}

In Fig. \ref{fig:M200m_sensitivity} we combine the analyses in Sec.~\ref{sec:probes} to present an overview of the sensitivity of different cosmological probes to halo populations as a function of their mass ($M_{\rm 200m}$) and redshift. We compare the sensitivity of the tSZ power spectrum at $\ell=500$ (\textit{Planck}; blue contour) and $\ell=3000$ (ACT and SPT; orange contour), the weak lensing two-point correlation function $\xi_{+}(\theta=4')$ in dark grey and $\xi_{-}(\theta=40')$ in light grey as measured from DES in its fourth tomographic bin, the eROSITA halo number counts and the DESI Legacy DR9 LRG catalog used for kSZ stacked profiles. For tSZ and weak lensing, the inner and outer contours correspond to 33\% and 66\% of the peak sensitivity (as in Figs.~\ref{fig:tsz_sensitivity} \& ~\ref{fig:WL_sensitivity}). For eROSITA, we show the 68\% and 95\% confidence interval of the (smoothed) halo number densities, and for kSZ a 2D hexagonal binning plot of the distribution of raw number counts. Grey lines show the median assembly histories of halos for a range of present day masses (denoted $M_{z=0}$), measured from the fiducial FLAMINGO simulation. 

The tSZ power spectrum is sensitive to $10^{15.1}\Mdot$ halos at $z\sim 0.1$ at low multipoles, and to $10^{14.7}\Mdot$ halos over the redshift range $z\in[0.2, 1]$, peaking at $z\sim 0.5$, at higher multipoles. The assembly history lines reveal that this is the \textit{same} population of halos, but observed at an earlier epoch in their assembly history. The median assembly histories are a faithful representation of the evolution of this population, as they experience a low number of major mergers from $z\sim1$ to $z\sim0$; only $\sim$18\% of them experience 1:5 mergers, $\sim$6\% 1:3, and $\sim$2\% 1:2, broadly consistent with previous work~\cite{Genel2009ApJ}.\footnote{\red{The merger ratios are computed by taking the ratio between the stellar masses of the infalling and reference objects at the pre-infall time.}} The weak lensing two-point correlation function of current surveys covers a broader range of halo masses ($10^{13.7-15.0}~\Mdot$) within a narrower range in redshift ($z\sim 0.1 - 0.4$), falling in the middle of the low-multipole and high-multipole tSZ contours. The eROSITA sample additionally overlaps with both tSZ (especially at higher multipoles), and with cosmic shear measurements.
This validates the typical use of X-ray calibration in weak lensing and tSZ analyses, after accounting for potential biases in the selection function of X-ray sources.
By contrast, the halos associated with the DESI LRG DR9 catalog for kSZ stacked profile measurements occupy a distinct mass and redshift range compared to all other probes: these are halos of masses $M_{\rm 200m}\sim 10^{12.5-13.5}~\Mdot$ at redshifts $0.4<z<1$. The mean halo mass of this sample is $M_{\rm 200m}\sim 10^{13.3}~\Mdot$; however, note that mean mass can shift by 0.1--0.2 dex depending on the exact stellar-to-halo mass prescription and the satellite fraction contamination.

In summary, we find two distinct halo populations: a higher-mass cluster population -- jointly probed by the tSZ power spectrum measured by \textit{Planck} at lower multipoles and ACT at higher multipoles, weak lensing measurements by DES, and X-ray measurements from eROSITA -- and a lower-mass group-scale population at redshifts $0.4<z<1$ associated with the kSZ profile measurements from the DESI LRG DR9 catalog.
This suggests that feedback constraints derived from any of former probes should be consistent, as they reflect the impact of feedback on the same underlying halo population; however, constraints drawn from the first set of probes do not dictate the expected impact of feedback in the kSZ measurements. 
The considered cosmological probes also differ in their sensitivity to radial scales around halos; we defer a detailed investigation of the impact of feedback at different radial scales to Sec.~\ref{sec:feedbackradial}.

\begin{figure*}
    \centering
    \includegraphics[width=\columnwidth]{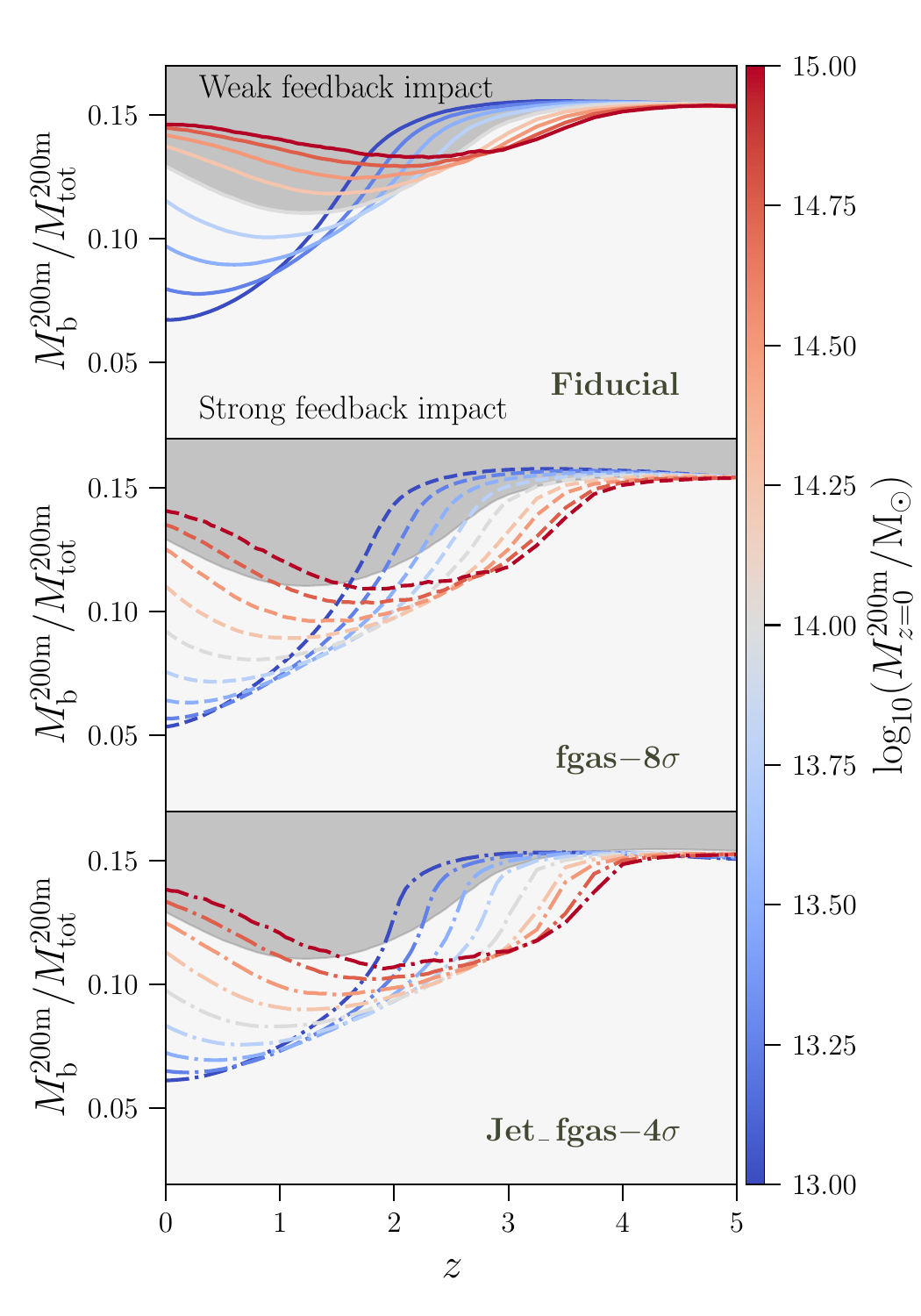}
    \includegraphics[width=\columnwidth]{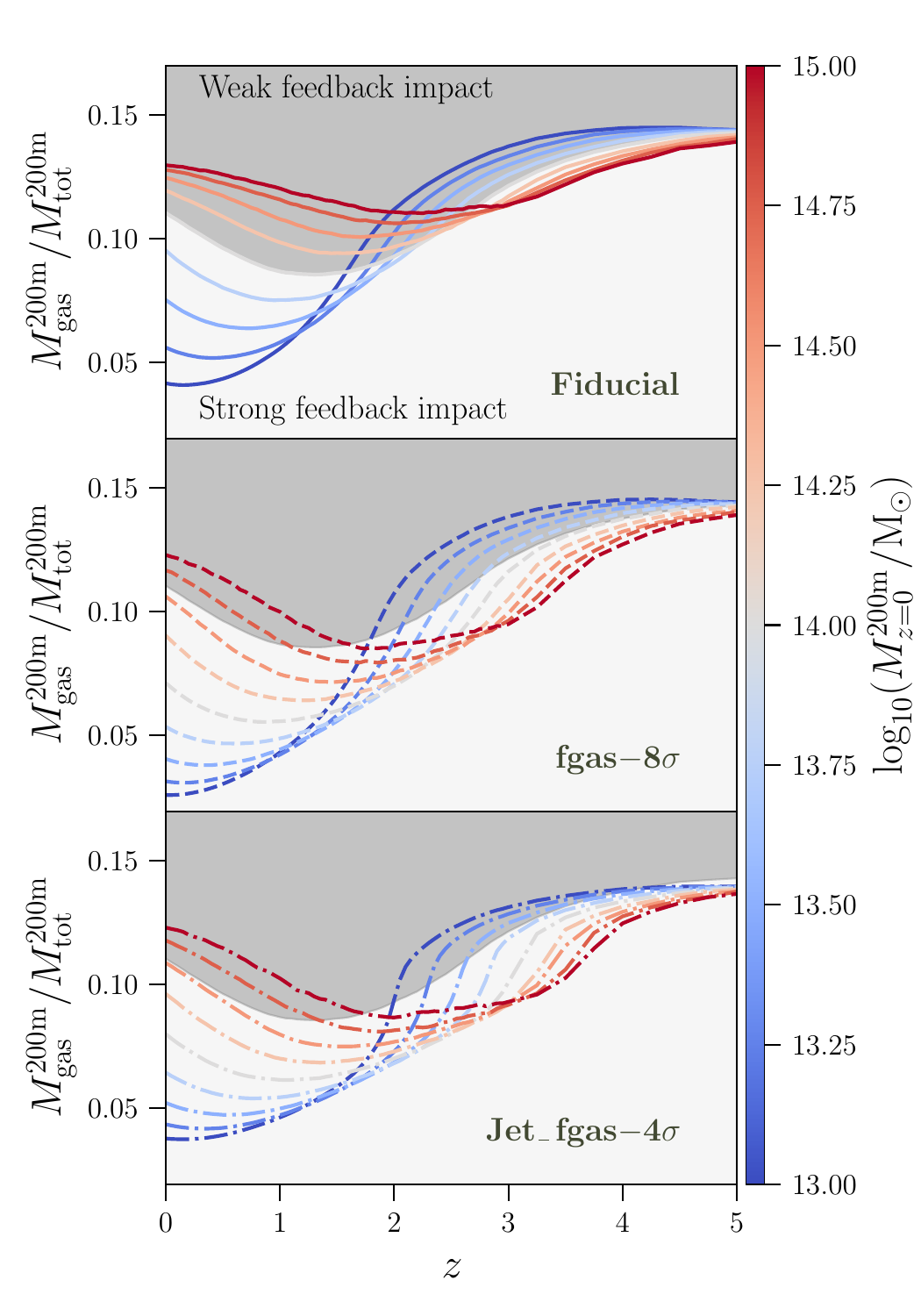}
    \caption{The fraction of baryon mass (left panels) and gas mass (right panels) relative to the total mass as a function of redshift for FLAMINGO halo populations with fixed present-day mass $M_{z=0}^{\rm 200m}$. The three panels, from top to bottom, show simulations with three different FLAMINGO feedback implementations: the fiducial, fgas$- 8\sigma$ and the jet-AGN implementation calibrated to fgas$- 4\sigma$. The grey region shows a regime of `weak' feedback impact: we define this demarcation line as corresponding to $M_{z=0}^{\rm 200m}>10^{14} M_{\odot}$, based on whether or not the halos' baryonic mass is significantly suppressed at $z=0$ in the fiducial simulation. This is also the regime of peak sensitivity for tSZ, weak lensing and eROSITA X-ray observations. While progenitors of these clusters were impacted by feedback at higher redshifts, these impacts on the mass fractions have largely dissipated by the time these halos are observed. The same grey region is repeated in the middle and lower panels to highlight the changes generated by the enhanced feedback variations. These variations have the expected effect of lowering average gas fractions across the population of groups and clusters, but have little impact on the highest mass clusters at $z<1$.}
    \label{fig:Mb_Mtot}
\end{figure*}

\begin{figure}
    \centering
    \includegraphics[width=\columnwidth]{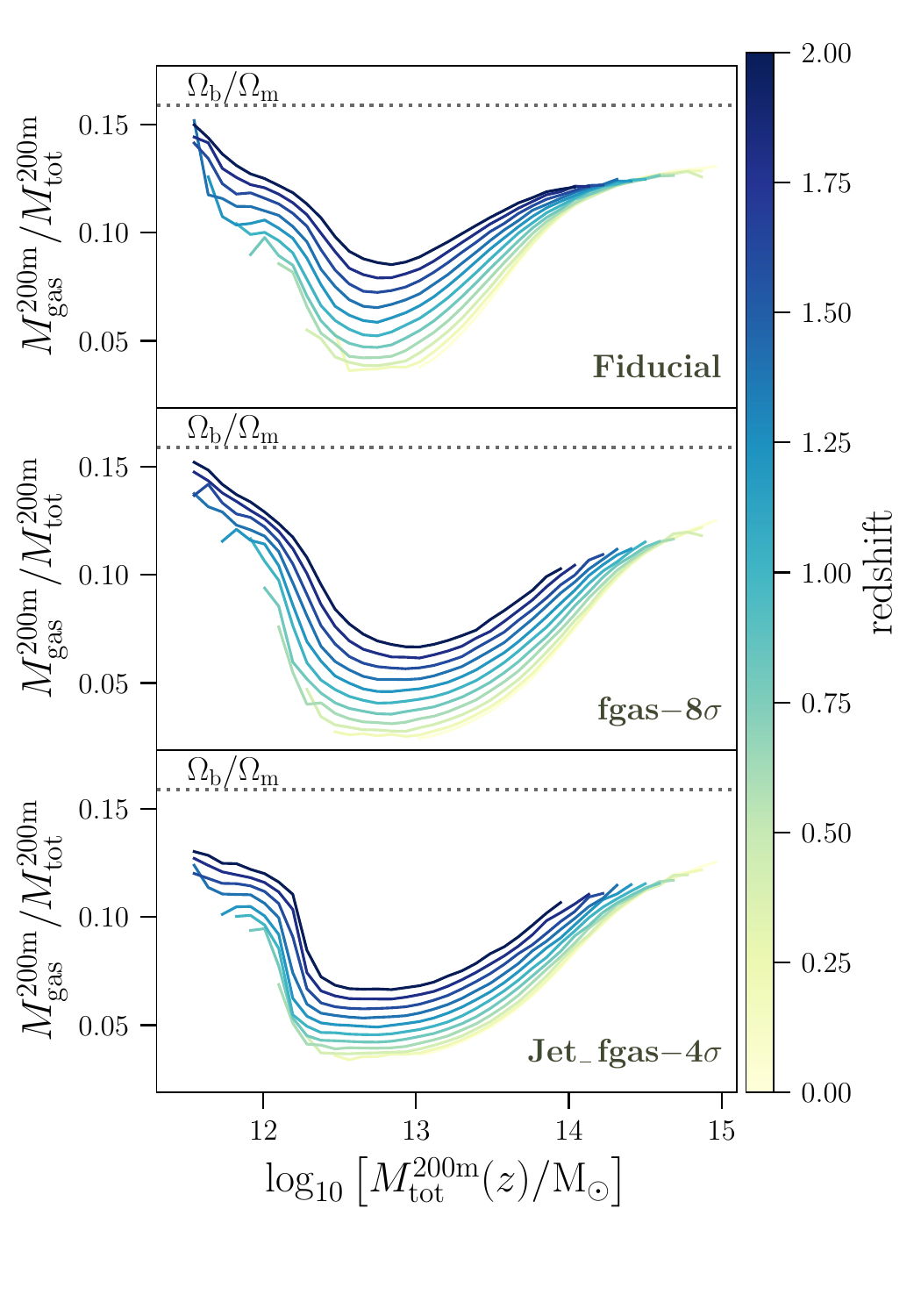}
    \caption{The fraction of gas mass relative to the total mass as a function of total halo mass at the specified redshift $z$. Different colored lines correspond to different redshifts; the three panels from top to bottom show respectively the results for the fiducial, fgas-$8\sigma$ and Jet\_fgas-$4\sigma$ simulations. The strongest suppression in the gas mass fraction occurs for halos of mass $M^{\rm 200m}_{\rm tot}\sim 10^{12.8}~\mathrm{M}_\odot$ across all redshifts. Different feedback prescriptions generate small shifts of 0.1-0.3 dex in this characteristic mass.}
    \label{fig:MgasMtot_z}
\end{figure}

The catalogs used in this work are broadly representative -- in terms of both mass and redshift -- of the wider set of galaxy catalogs and X-ray measurements available in the literature. For example, the stellar masses and redshifts of the DESI LRG catalog used here, as measured by \citet{Zhou_2023}, closely match those of the BOSS CMASS and LOWZ catalogs, which have been used for previous kSZ stacked measurements \cite{Schaan2021}. \citet{Maraston2013} find a narrow stellar mass distribution for the BOSS galaxies in the range $\log_{10} M_\star\sim [11, 12]~\Mdot$ at redshifts $0.2<z<0.6$, peaking at $\log_{10} M_\star \sim 11.3$; these values are consistent with those of the DESI LRGs reported here (see Appendix~\ref{app:SMHM}). \citet{mccarthy2024flamingocombiningkineticsz} use an alternative method based on galaxy-galaxy lensing to measure the mass of BOSS CMASS and LOWZ galaxies, reporting minimum stellar masses $\log_{10} M_\star\sim 11.3$ for LOWZ and $\log_{10} M_\star\sim 11.2$ for CMASS; although slightly higher than those of the DESI LRGs, these values remain in qualitative agreement.

Similarly for X-ray measurements, there exist alternative catalogs than those presented in this work from eROSITA, including those from the XMM-XXL survey from the Hyper Suprime-Cam (HSC) Subaru Strategic Program survey. The latter provides mass measurements of 136 X-ray galaxy groups and clusters using weak lensing \cite{Umetsu_2020, Akino_2022}. The masses cover a wide range of values $M_{\rm 200}\sim 10^{13.3} -10^{15.3}~\Mdot$ at an average redshift $z\sim0.3$; these again are in qualitative agreement with our reported values.

\subsection{Impact of baryonic feedback during halo assembly}
\label{sec:feedback}
Having distinguished between two halo populations -- each characterized by different masses and redshifts and probed by different cosmological observables -- we now investigate the impact of baryonic feedback on these populations. The physical effect responsible for the suppression of the matter power spectrum is the re-distribution of baryons within halos mainly due to AGN feedback. Therefore, we investigate the baryonic (and gas) mass fraction within halos over time, allowing us to quantify and disentangle the effects of baryonic feedback over the halos' evolutionary histories.  

We show the impact of baryonic feedback on different halo populations in Fig.~\ref{fig:Mb_Mtot}. We plot the fraction of baryon mass relative to the total mass (left panel) and the fraction of gas mass relative to the total mass (right panel) for halo populations with fixed present-day mass. The three panels, from top to bottom, show simulations with three different feedback implementations: the fiducial case, fgas$- 8\sigma$ and the jet-AGN implementation calibrated to fgas$- 4\sigma$. We qualitatively split the top-panel into `weak' and `strong' feedback impact regimes based on the assembly history of an intermediate mass halo, which roughly separates halos whose baryonic mass is increasingly suppressed due to AGN feedback (light grey region) from those that are not (darker grey region).

The most massive halo population (dark red) is that probed jointly by tSZ, weak lensing and eROSITA measurements at redshifts $z<1$. In this observable redshift range, this population lies within the ``weak feedback impact'' regime for all three feedback implementations. Feedback affects high-mass clusters at early times ($z\sim 2 - 4$); however, as these clusters assemble more mass, feedback becomes inefficient ($z<2$), and the clusters restore their pre-feedback mass. Part of this effect comes from the fact that, although baryons are expelled at early times, these do not travel far enough to escape the gravitational potential wells of the clusters and therefore are re-accreted at later times. In addition to purely gravitational re-accretion, feedback from surrounding objects blows gas into the Lagrangian region destined to become the cluster \cite{Mitchell_2021}.

The lower-mass populations (dark blue), which are sensitive to kSZ stacked profile measurements, lie instead within the `strong feedback impact' regime. These halos are largely unaffected by feedback at early times until $z\sim 2$, meaning that the fraction of baryonic (or gas) mass remains constant during this time. After $z\sim 2$, feedback becomes efficient at expelling gas beyond $r_{\rm 200m}$ leading to a suppression of the baryonic (or gas) mass within those scales.

\begin{figure*}
    \centering
    \includegraphics[width=0.95\columnwidth]{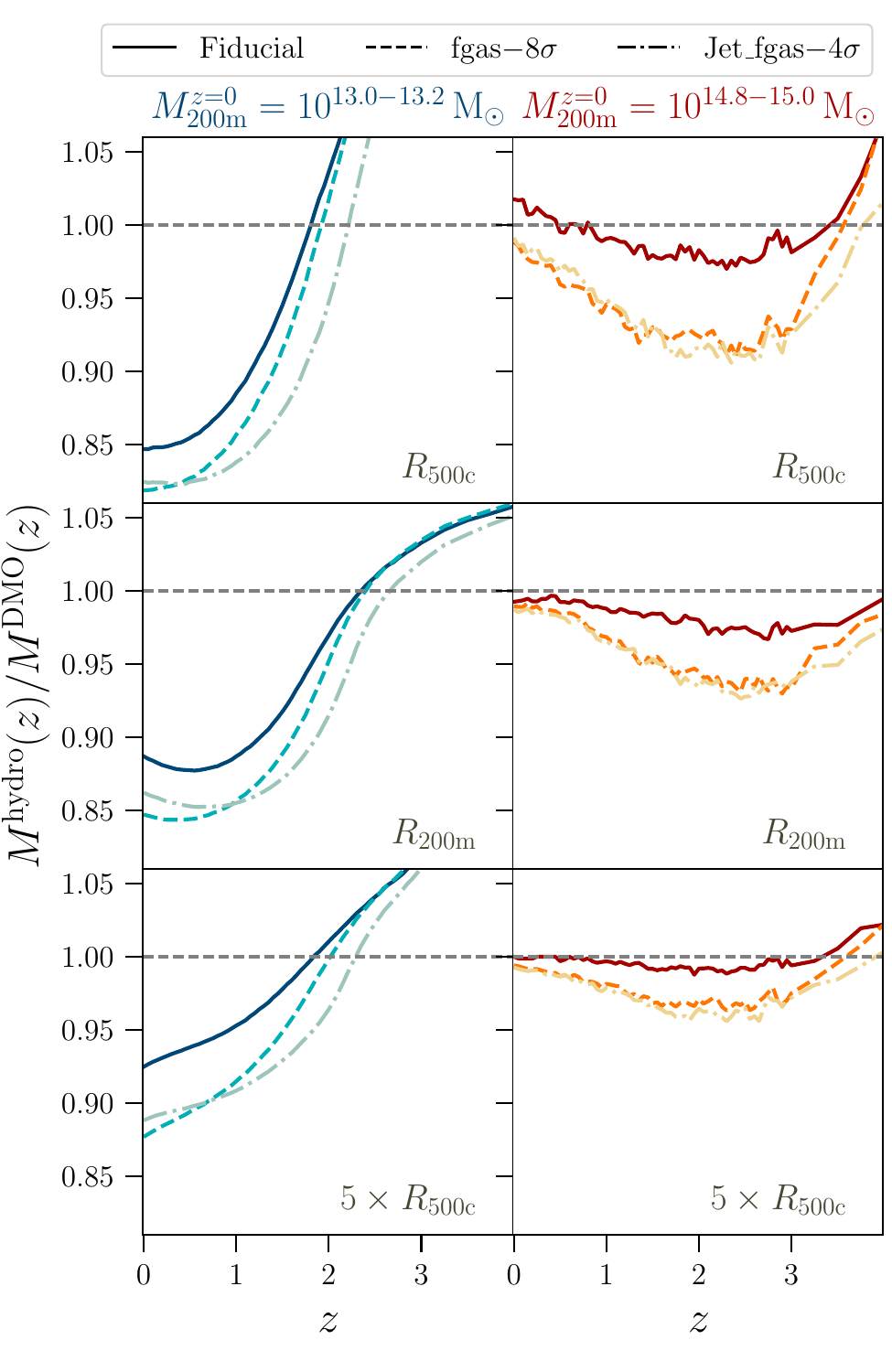}
    \hspace{1cm}
    \includegraphics[width=0.95\columnwidth]{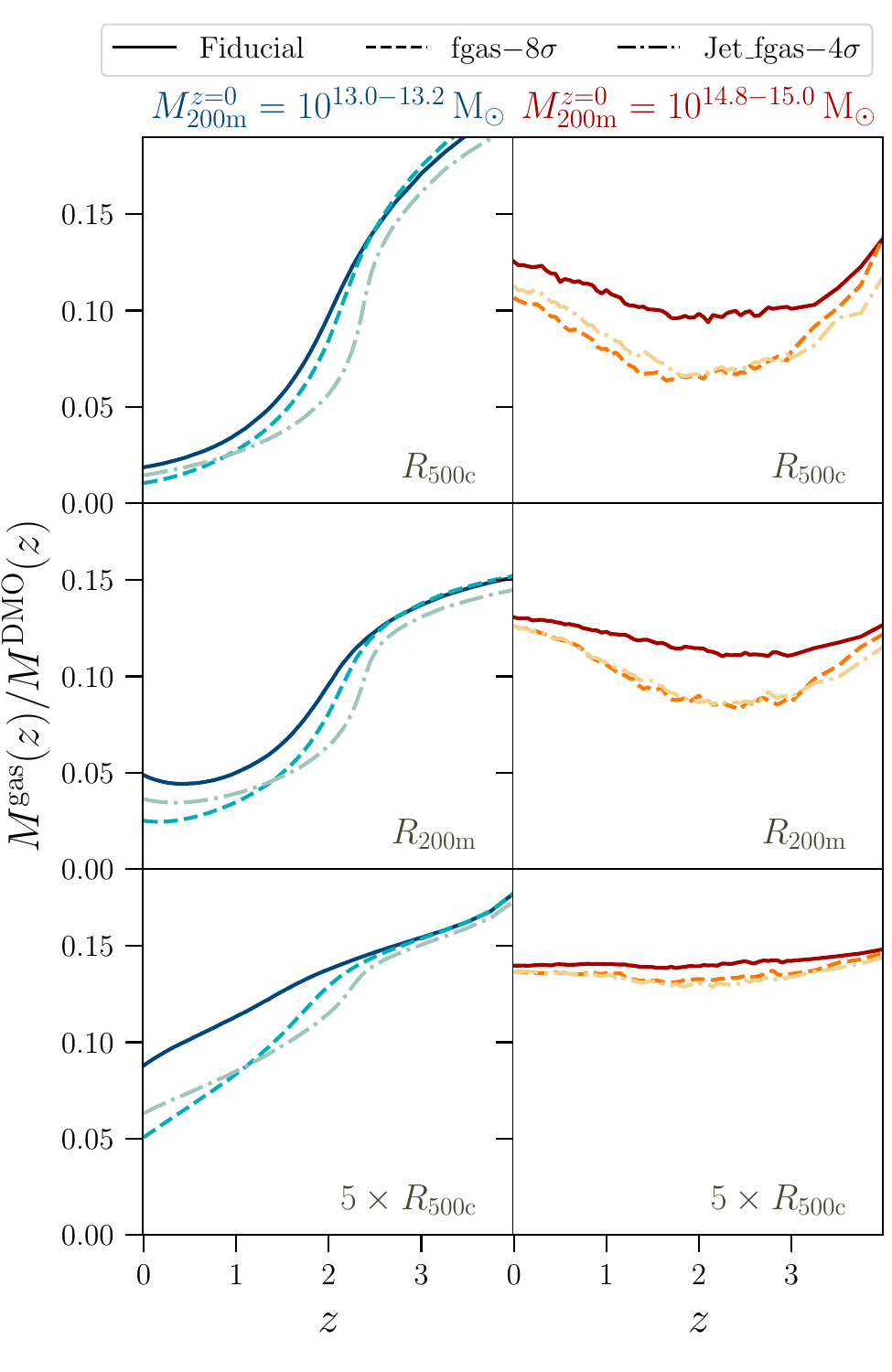}
    \caption{\textit{Left panel:} Median of the ratio between the total mass accretion histories of a halo in a hydrodynamical simulation and its matched counterpart in the gravity-only simulation. The columns respectively show mass scales relevant to kSZ stacks ($M_{\rm 200m}^{z=0}=10^{13.0 - 13.2}\, \Mdot$; left column) and tSZ, X-ray and weak lensing power spectra ($M_{\rm 200m}^{z=0}=10^{14.8 - 15.}\, \Mdot$; right column). Different line styles correspond to differing feedback implementations. The three rows adopt differing mass definitions as proxies for different radial scales, moving from inner to outer regions from top to bottom respectively. Different observables measure the distribution of baryons or gas out to different spatial scales: kSZ is sensitive to the distribution of baryons in group-scale halos even out to their outer regions (bottom-left), tSZ and lensing reach out to $\sim \mathrm{R_{\rm 200m}}$ (middle-right) in higher-mass clusters, while X-rays probe the inner regions of the same high-mass cluster population (top-right). \textit{Right panel:} Same as the left panel, except the y-axis shows the ratio between the gas mass of the halos in the hydrodynamical simulations relative to the total mass of their respective counterpart halos in the gravity-only simulations. 
    }
    \label{fig:MAH_ratio}
\end{figure*}

The grey regions denoting weak (or strong) feedback impact are repeated in the middle and lower panels to contextualize the changes generated by the enhanced feedback variations. As expected, both variations lower average gas and baryon fractions across the population of groups and
clusters, but retain the same overall trend. In the case of the jet-like AGN (bottom panels), the ejection of gas from the halo happens at a faster rate before $z\sim2$, thus leading to a steeper change in gas mass relative to total mass compared to the case of enhanced feedback with thermal AGN (middle panels). Feedback processes have relatively little impact on the highest mass clusters at $z < 1$ even for more extreme feedback models.

The comparison between the left and right panels highlights differences in the impact of feedback on the total baryon mass and on the gas mass alone. This distinction is relevant due to the fact that weak lensing is sensitive to the total mass in clusters, while tSZ, kSZ and X-ray observations probe the gas alone. We find that the baryonic mass of $M_{\rm 200m}\sim10^{13.1}~\Mdot$ halos -- i.e., those probed by kSZ measurements -- is $\sim$ 7\% of the total mass at $z=0$ in the fiducial case, but gets suppressed further to 5\% in the `fgas$- 8\sigma$' model. The gas mass of the same halo mass population instead reduces from 4\% to 2\% at $z=0$ for the two models, which implies a bigger change in gas than for the baryons (50\% vs. 30\% shift respectively). For high-mass clusters -- probed by tSZ, lensing and X-ray observations -- the impact of feedback is negligible for both baryonic and gas mass (a $\sim$5\% shift).

We conclude that baryonic feedback does not manifest in the same way for halos detected through X-ray, lensing and tSZ measurements, and those through the kSZ effect: it is possible for the latter to be compatible with enhanced-feedback scenarios, while high-mass clusters remain compatible with more modest feedback strengths as inferred from X-ray observations. This explains the apparent tension \cite{Bigwood2024} in estimates of gas mass fractions when adopting kSZ as opposed to X-ray observations. Specifically, the baryon fraction in high-mass clusters approaches the cosmic mean, while in galaxy groups the fraction is about half as large \cite{Eckert2021}, consistent with the picture in Fig. 4.

\subsection{Impact of baryonic feedback at fixed time}
\label{sec:feedbackfixedtime}
The dependence on redshift in the above results arises solely due to the changing mass of a given population over time. 
When seen as a function of instantaneous mass, feedback is most efficient at expelling gas outside halos of mass $M_{\rm 200m}\sim10^{12.8}~\Mdot$, independent of redshift. 
In other words, feedback is most efficient at that mass scale, regardless at what redshift a given halo reaches that mass during its evolution.
This is demonstrated in Fig.~\ref{fig:MgasMtot_z}, where we show the fraction of gas relative to the total halo mass as a function of the latter for different redshifts. We show the three simulations with different feedback calibrations (fiducial, fgas-$8\sigma$, and Jet\_fgas-4$\sigma$) in the three panels from top to bottom, respectively. We find that largest suppression in gas mass relative to total mass occurs at $M_{\rm 200m}\sim10^{12.8}~\Mdot$ for all redshifts. This result holds across the three feedback variations, with small variations in the maximum suppression mass scale within the range $M_{\rm 200m}\sim10^{12.5-13.0}~\Mdot$. The rate at which feedback becomes less efficient as halos accrete more mass above $10^{12.8}~\Mdot$ depends on the specific feedback implementation. By the time halos reach $M_{\rm 200m}\sim10^{14}~\Mdot$, again independent of the redshift at which this happens, feedback ceases to be effective. It would be interesting to check the extent to which these findings generalize to different simulation suites. From semi-empirical models, it is known that the peak of star formation efficiency does not evolve strongly with cosmic time \cite{Behroozi2013ApJ}. However, this peak occurs at a halo mass of $\approx 10^{12}\, \Mdot$, approximately 1 dex lower than the relevant scale found in this work for the peak AGN impact.

\subsection{Radial dependence of baryonic feedback}
\label{sec:feedbackradial}
\begin{figure*}
    \centering
    \includegraphics[width=0.94\textwidth
    ]{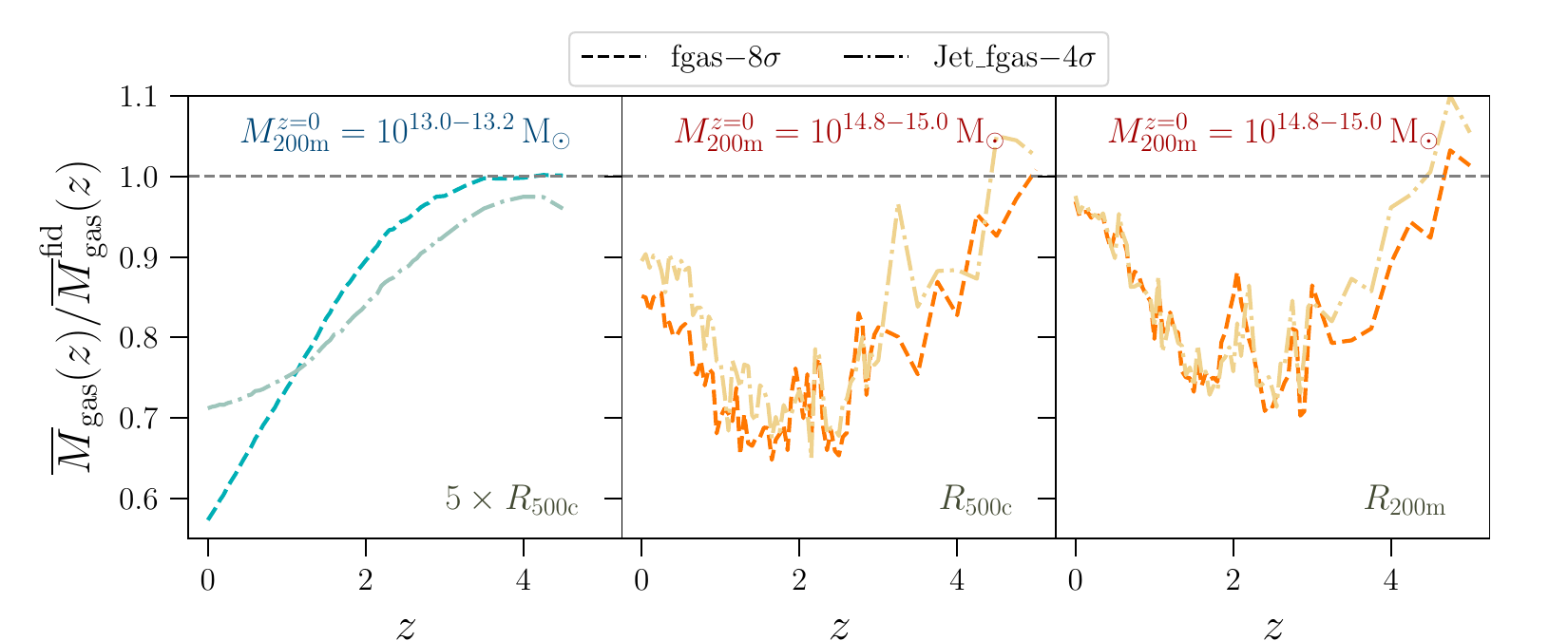}
    \caption{A more detailed view of the differences in the median gas mass histories for different feedback implementations relative to the fiducial case. We focus on the mass and radial scales of halos relevant to different observables: kSZ is sensitive to lower-mass groups out to their outer regions ($M_{\rm 200m}^{z=0} = 10^{13.0 - 13.2}\Mdot$ and $r \sim 5\times R_{\rm 500c}$; left panel), X-ray measurements to the inner region of higher-mass clusters ($M_{\rm 200m}^{z=0} = 10^{14.8 - 15.0}\Mdot$ and $r \sim R_{\rm 500c}$; middle panel), and tSZ and weak lensing to the virial region of the same higher-mass clusters ($M_{\rm 200m}^{z=0} = 10^{14.8 - 15.0}\Mdot$ and $r \sim R_{\rm 200m}$; right panel). At $z\ll1$, feedback variations within FLAMINGO allow for a $\sim 15\%$ suppression in gas mass in the inner regions ($R_{500c}$, middle panel) but only a $\sim 5\%$ suppression in the outer regions ($R_{200m}$, right panel). The two enhanced feedback variations have approximately the same impact on the gas relative to the fiducial case.}
    \label{fig:Mgas_ratio}
\end{figure*}

We next examine how different feedback prescriptions affect the gas content within halos across various radial scales.
We return to selecting populations with a fixed $z=0$ mass, and  focus on representative cases of interest: lower-mass groups ($M_{\rm 200m}^{z=0} \simeq 10^{13.1}\,\Mdot$ in the DMO simulation) as probed by stacked kSZ profiles, and higher-mass clusters ($M_{\rm 200m}^{z=0} \simeq 10^{14.9}\,\Mdot$ in the DMO simulation) as probed by tSZ, weak lensing and X-ray measurements. 

Fig.~\ref{fig:MAH_ratio} shows the median ratio of the mass accretion histories between halos in the hydrodynamical simulations and their counterparts in the gravity-only simulations. The two chosen populations are shown in the left and right columns respectively, along with the effects of three different feedback implementations (different lines in each panel). The three rows correspond to different mass definitions, serving as proxies for radial scales that range from the inner to the outer halo regions (top to bottom panels). X-ray measurements primarily measure the gas fraction within the inner region of high-mass clusters (top-right panel) \cite{Vikhlinin2005ApJ}; tSZ and weak lensing probe the gas and baryonic content respectively out to $~R_{\rm 200m}$ of the same high-mass clusters (middle-right panel) \cite{Planck2016tSZ, Amon_2022}; kSZ instead probes the amount of gas in the outskirts of group-sized halos (bottom-left corner) \cite{Schaan2021}.

We find that high-mass clusters are largely insensitive to feedback even for enhanced feedback models, as their mass $M_{\rm hydro}\simeq M_{\rm DMO}$ for $z<1$; this is true at all radial scales and even in the inner regions probed by X-ray measurements where differences are only a few percent. Moreover, this appears true for both the total baryonic content (left panels) and the gas mass alone (right panels). At higher redshift ($z\sim2$) however, baryonic feedback is able to suppress the total mass of halos by $\sim$10\%. On the other hand, feedback has a significant effect ($\sim 5-25\%$) on low-mass halos depending on the strength of feedback; this impact remains significant even out to the outermost regions of halos which kSZ stacked measurements are sensitive to. 

The kSZ is sensitive only to the gas mass, rather than the total mass. As such, changes in the feedback have a particularly strong effect on kSZ predictions, far stronger than the total mass profile would suggest. While this can be inferred by comparing the lower-left panel of the left group and the right group in Fig.~\ref{fig:MAH_ratio}, it is even clearer in Fig. \ref{fig:Mgas_ratio}, which shows the ratio of the median gas mass histories in the enhanced-feedback models relative to the fiducial case. The left panel in blue shows the impact of feedback in the outer region of the halos ($5\times R_{\rm 500c}$), directly illustrating the strong sensitivity of kSZ to feedback adjustments. The suppression in gas mass due to feedback variations can lead to $\sim40$\% larger suppression than the fiducial case.

The middle and right panels of Fig.~\ref{fig:Mgas_ratio} illustrate the effect of varying feedback on gas mass in the representative population of X-ray and tSZ clusters respectively. (Due to these observables' dependence on temperature as well as density, the total mass also plays an indirect role, but here we continue to focus on gas mass for simplicity.) We find that, at $z < 1$, the impact of model variations on the gas content relative to the fiducial case within high-mass clusters can also be significant. 
Our findings are consistent with the results of \citet{McCarthy2023} who show that feedback has little impact on the tSZ power spectrum at lower multipoles (lower redshifts) probed by \textit{Planck}, but has some effect on higher multipoles corresponding to higher redshift clusters probed by ACT and SPT.
The largest feedback variations within FLAMINGO allow for a $\sim$ 15--30\% suppression in gas mass in the inner regions of groups ($R_{\rm 500c}$, middle
panel) and a $\sim$ 5--20\% suppression in the virial regions of massive clusters ($R_{\rm 200m}$, right panel). In all cases, the two enhanced feedback variations examined here have approximately the same impact on the gas relative to the fiducial case.

\section{Discussion and Conclusions}
\label{sec:conclusions}
We have examined the impact of baryonic feedback on cosmological observables, highlighting how this should be seen primarily in the context of halo assembly ($M(z)$) rather than as a function of wavenumber ($P(k,z)$). 
We have investigated the imprint of feedback on different cosmological observables in terms of the underlying halo populations (characterized by their mass, redshift and radius). The sensitivity with respect to halo mass and redshift has been computed using a combination of analytic calculations based on the halo model and observational data catalogs (Fig.~\ref{fig:M200m_sensitivity}), while the impact of feedback in terms of halo assembly histories has been assessed using the FLAMINGO simulations. The latter has also been used to measure the radial dependence of the imprints of feedback on these halo populations (Figs.~\ref{fig:MAH_ratio} \& \ref{fig:Mgas_ratio}). 

The effect of FLAMINGO feedback is a function of mass, radius and redshift. Our results show that feedback most efficiently redistributes baryons when halos reach a mass of $M_{\rm 200m}\sim 10^{12.8}~\Mdot$, independent of redshift (Fig.~\ref{fig:MgasMtot_z}). As the halo mass $M(z)$ grows beyond this threshold, the gas is reaccreted, so that high-mass clusters ($M_{\rm 200m}\simeq 10^{15}~\Mdot$) are barely sensitive even to strongly enhanced feedback (Fig.~\ref{fig:Mb_Mtot}). Thermal SZ power spectra, X-ray clusters, and cosmic shear observables all have maximum sensitivity to feedback at this high-mass clusters scale. On the other hand, the halos probed by the kSZ effect generally cover a lower mass range at intermediate redshifts ($M_{\rm 200m}\sim 10^{13}~\Mdot$, $z \sim 0.5 - 1$) compared to the other probes. This gives a natural explanation for why stacked kSZ observations around galaxy groups can be explained with a modest strengthening of AGN feedback \cite{mccarthy2024flamingocombiningkineticsz} or the inclusion of cosmic rays \cite{Quataert25}, while tSZ power spectrum observations are much harder to explain using feedback variations \cite{McCarthy2023}. Our conclusions are in agreement with new constraints from eROSITA \cite{Popesso2024} which indicate that the fiducial FLAMINGO feedback implementation is compatible with estimated gas masses for $M_{500} \le 10^{13.5} \, \Mdot$ and $M_{500} \ge 10^{14.5} \, \Mdot$ but too high in the intermediate range (see their figure 6). This is the range where gas mass is most strongly altered by the shift to the fgas$-8\sigma$ calibration (our Fig.~\ref{fig:Mb_Mtot}).

If the discrepancy between tSZ power predictions and observations at $\ell>1000$ is related to feedback, it must correspond to a long-lived reduction in the gas density without raising the electron temperature. 
One way to achieve this via thermal feedback is by ejecting the gas sufficiently far out in the outskirts of halos, thus also leading to adiabatic cooling (see e.g., Refs.~\cite{McCarthy_2011,Altamura2025}).
Non-thermal sources of pressure may also contribute to these effects; for example, additional sources of turbulence to those attributed to accreting material \cite{Shi2014}, magnetic fields and cosmic rays, although \citet{Quataert25} caution that the latter do not easily contribute at such large halo masses. Alternatively, a greater fraction of diffuse baryons in the outskirts of clusters could be locked up in stars than currently expected (see e.g., recent observational findings from the intracluster light at relevant redshifts \cite{JooICL2023}). In this scenario, these massive clusters would have little overall baryonic mass suppression. A potentially powerful technique to differentiate between feedback mechanisms is to combine measurements that probe the gas within the same halo population (e.g., galaxy-galaxy lensing and kSZ profiles of the same stacked population \cite{mccarthy2024flamingocombiningkineticsz}). We defer investigating this scenario to future work. 

The total baryonic mass of halos has been raised as an important consideration in resolving apparent tensions in cosmological data. In particular, the $S_8$ parameter characterizes the amplitude of matter density fluctuations in the Universe, and is consistently reported to be lower when constrained through lensing than CMB observations suggest  \cite{2021A&A...646A.140H,2022PhRvD.105b3520A}. While baryonic feedback has been raised by several authors as a route to resolving this discrepancy \cite[e.g.,][]{AmonEfstathiou2022, Preston2023}, others have argued that it is insufficiently strong to resolve the tension on its own \cite[e.g.,][]{McCarthy_2018, McCarthy2023}. Meanwhile, more recent lensing results appear anyway to be moving towards agreement with {\it Planck} results \cite{2025arXiv250319442S,Wright2025, DESKiDS2023}, without making any significant changes in accounting for the baryonic feedback.

The long-standing disagreement between the measured tSZ power spectrum from \textit{Planck} and predictions from hydrodynamical simulations at $\ell<1000$ has also been framed in part as a cosmological tension, challenging to account for with feedback but potentially resolvable with a lower $S_8$ value \cite{mccarthy2024flamingocombiningkineticsz}. However, a recent re-analysis of \textit{Planck} data led to a shift in the measurements and an increase in the associated uncertainties, bringing the measurements and theoretical predictions at low-$\ell$ into closer agreement \cite{efstathiou2025}, again without the need for any extreme feedback or changes to $S_8$.

Given these latest results from lensing and tSZ, the status of the $S_8$ tension is currently unclear. However it seems unarguable that a self-consistent accounting of feedback in cosmology must be able to describe both weak lensing and tSZ constraints simultaneously. In this paper, we have shown that the sensitivity of lensing to the relevant one-halo terms after DES scale cuts probes the same halo population as is responsible for tSZ power. If one instead applies KiDS scale cuts, the one-halo term starts to probe a lower mass range which does not overlap significantly with halo populations probed either by tSZ or kSZ. 

If one is interested in recovering cosmological parameters from lensing independently of these feedback effects, our results confirm that scale-cuts similar to those adopted by DES are expected to be effective in minimizing the one-halo contribution. The strong localization in halo mass and redshift of the remaining one-halo contribution suggests that explicit nulling methods for shaping the lensing kernel \cite{Piccirilli2025, Taylor_2021, Bernardeau_2014} will perform even better. Moreover, a simple strategy of combining scale cuts with masking the small number of very high mass clusters could similarly prove effective. 

\section*{Acknowledgments}
We thank Alexandra Amon, Boris Bolliet, Marcus Br{\"u}ggen, Esra Bulbul, Vittorio Ghirardini, Boryana Hadzhiyska, Eiichiro Komatsu, Elizabeth Krause, Ian McCarthy, Volker Springel, Tom Theuns, Chun-Hao To, Cora Uhlemann and Rongpu Zhou for useful discussions. 
We thank the Virgo Consortium, in particular Carlos Frenk, Adrian Jenkins and Baojiu Li, for granting us access to the COSMA cluster.
LLS acknowledges support by the Deutsche Forschungsgemeinschaft (DFG, German Research Foundation) under Germany’s Excellence Strategy – EXC 2121 ``Quantum Universe'' – 390833306. HVP and AH have been supported by funding from the European Research Council (ERC) under the European Union's Horizon 2020 research and innovation programmes (grant agreement no. 101018897 CosmicExplorer). HVP was additionally supported by the G\"{o}ran Gustafsson Foundation for Research in Natural Sciences and Medicine. HVP acknowledges the hospitality of the Aspen Center for Physics, which is supported by National Science Foundation grant PHY-1607611. AP has been supported by funding from the European Research Council under the European Union's Horizon 2020 research and innovation programmes (grant agreement no. 818085 GMGalaxies). 
This work used the DiRAC@Durham facility managed by the Institute for Computational Cosmology on behalf of the STFC DiRAC HPC Facility (www.dirac.ac.uk). The equipment was funded by BEIS capital funding via STFC capital grants ST/K00042X/1, ST/P002293/1, ST/R002371/1 and ST/S002502/1, Durham University and STFC operations grant ST/R000832/1. DiRAC is part of the National e-Infrastructure.

\section*{Author contributions}
We describe the different author contributions using the CRediT\footnote{\href{https://credit.niso.org/}{https://credit.niso.org}} (Contribution Roles Taxonomy) system.
\textbf{L.L.-S.:} Conceptualization (supporting); Formal analysis; Investigation; Methodology; Validation; Visualization; Writing - original draft, review \& editing.
\textbf{H.V.P.:} Conceptualization (lead); Methodology; Investigation; Validation; Visualization; Writing - original draft, review \& editing.
\textbf{A.P.:} Conceptualization (supporting); Methodology; Investigation; Writing - original draft, review \& editing.
\textbf{A.H.:} Formal analysis; Writing - original draft, review \& editing.
\textbf{J.S.:} Project administration (FLAMINGO); Data curation; Resources; Writing - Review \& Editing.
\textbf{M.S.:} Data curation; Resources.
\textbf{J.C.H.:} Data curation; Resources.
\textbf{R.J.M:} Data curation; Resources.
\textbf{W.E.:} Data curation.


\appendix
\section{Stellar mass-to-halo mass relation}
\label{app:SMHM}
\begin{figure}
    \centering
    \includegraphics[width=\columnwidth]{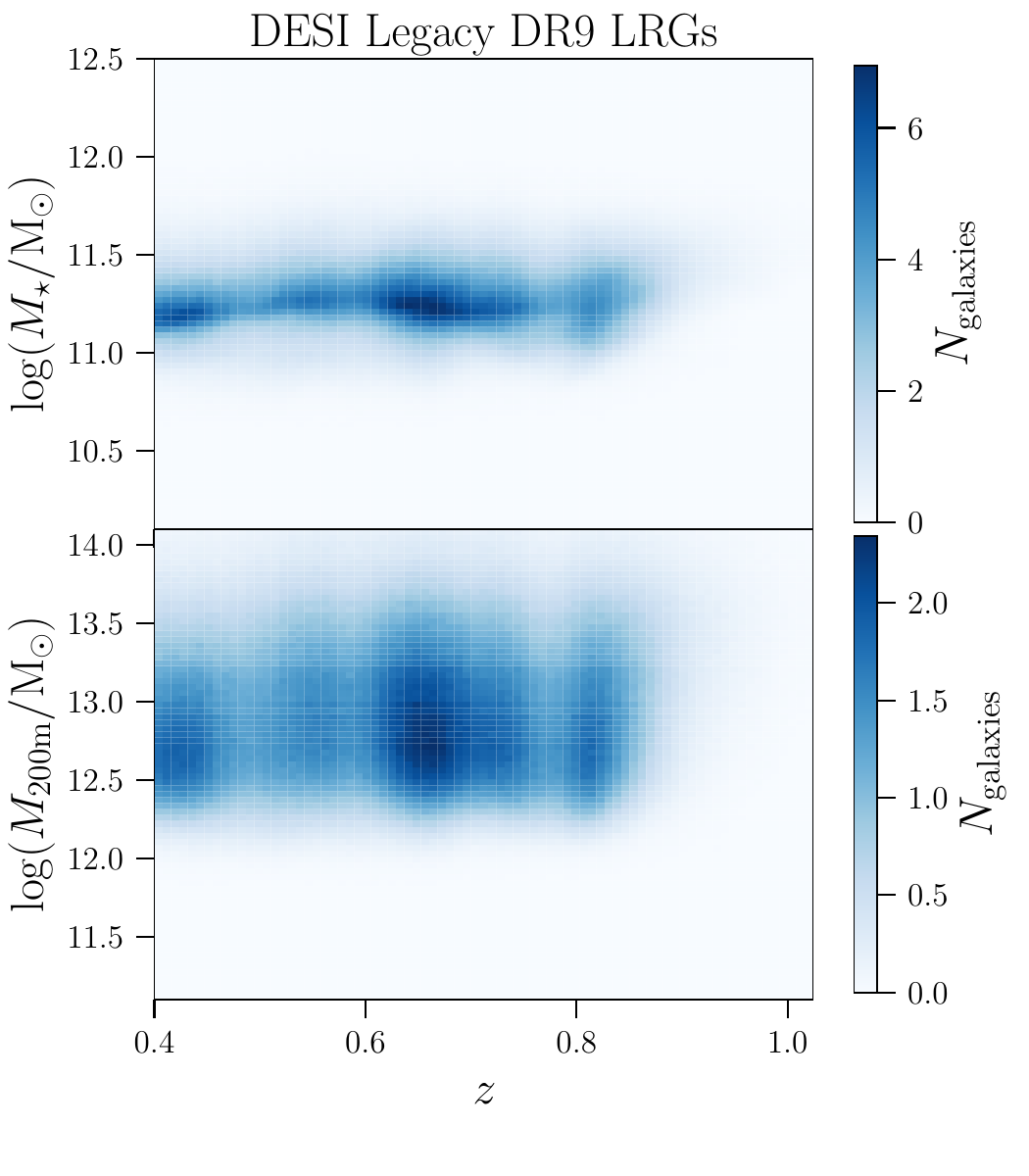}
    \caption{2D histogram of the stellar masses (top panel) and halo masses (bottom panel) of the DESI Legacy DR9 LRGs. The halo masses were obtained using a SMHM relation fitted to the stellar masses and halo masses in the FLAMINGO fiducial simulation, which also accounts for the scatter in the halo mass at fixed stellar mass.}
\label{fig:SMHM_app}
\end{figure}

To establish a connection between the stellar mass of a galaxy $M_\star$ at a given redshift $z$ and the mass of its associated dark matter halo $M_{\rm 200m}$, we derive an empirical stellar-mass-to-halo-mass relation from the fiducial FLAMINGO simulation at a representative redshift $z=0.7$.
We consider central galaxies only; see Sec.~\ref{sec:ksz} for justification of this choice and further discussion. We split the stellar masses of the galaxies into thin, evenly-spaced bins, and sample from the distribution of halo masses within each bin; this allows us to assign to any given stellar mass $M_\star$ a corresponding $M_{\rm 200m}$ which reflects the stellar-to-halo mass relation in FLAMINGO, including the scatter in halo mass at fixed stellar mass. 

We applied this procedure to the stellar mass values of the DESI Legacy DR9 LRGs sample. We show in Fig.~\ref{fig:SMHM_app} the 2D histogram of stellar masses as a function of redshift in the top panel, and the corresponding halo masses in the bottom panel.

\begin{figure*}
    \centering
    \includegraphics[width=0.49\textwidth]{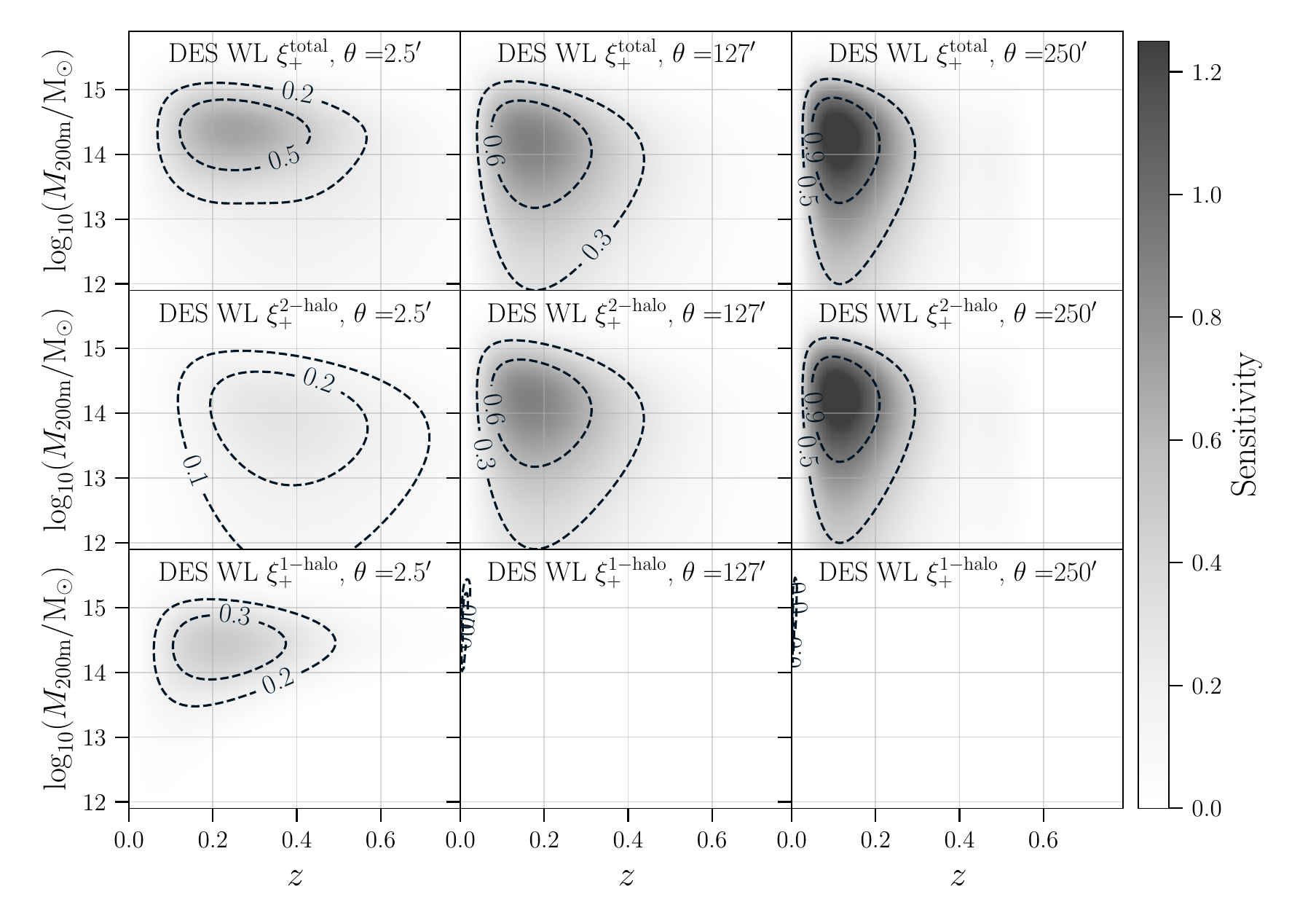}
    \includegraphics[width=0.49\textwidth]{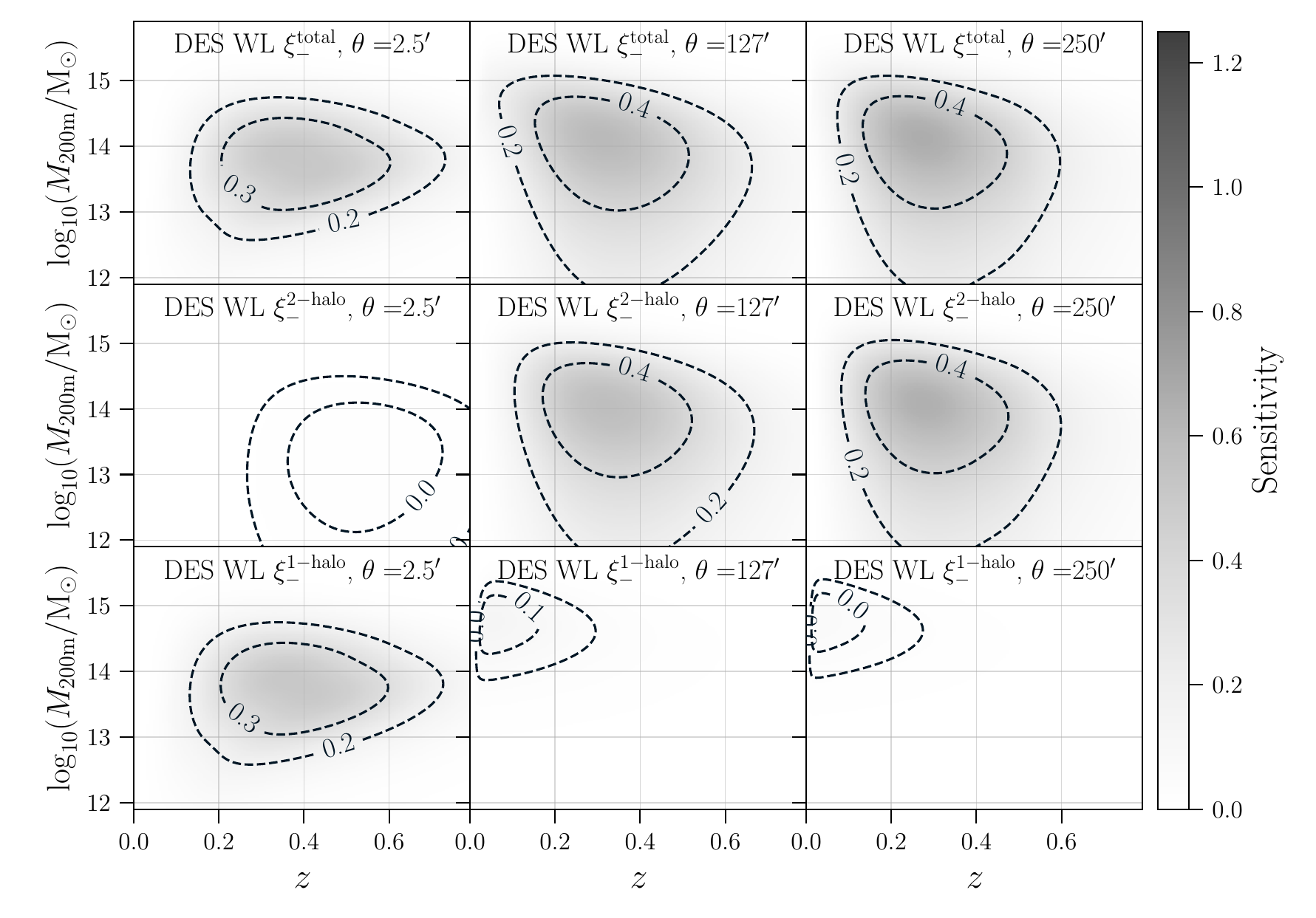}
    \includegraphics[width=0.49\textwidth]{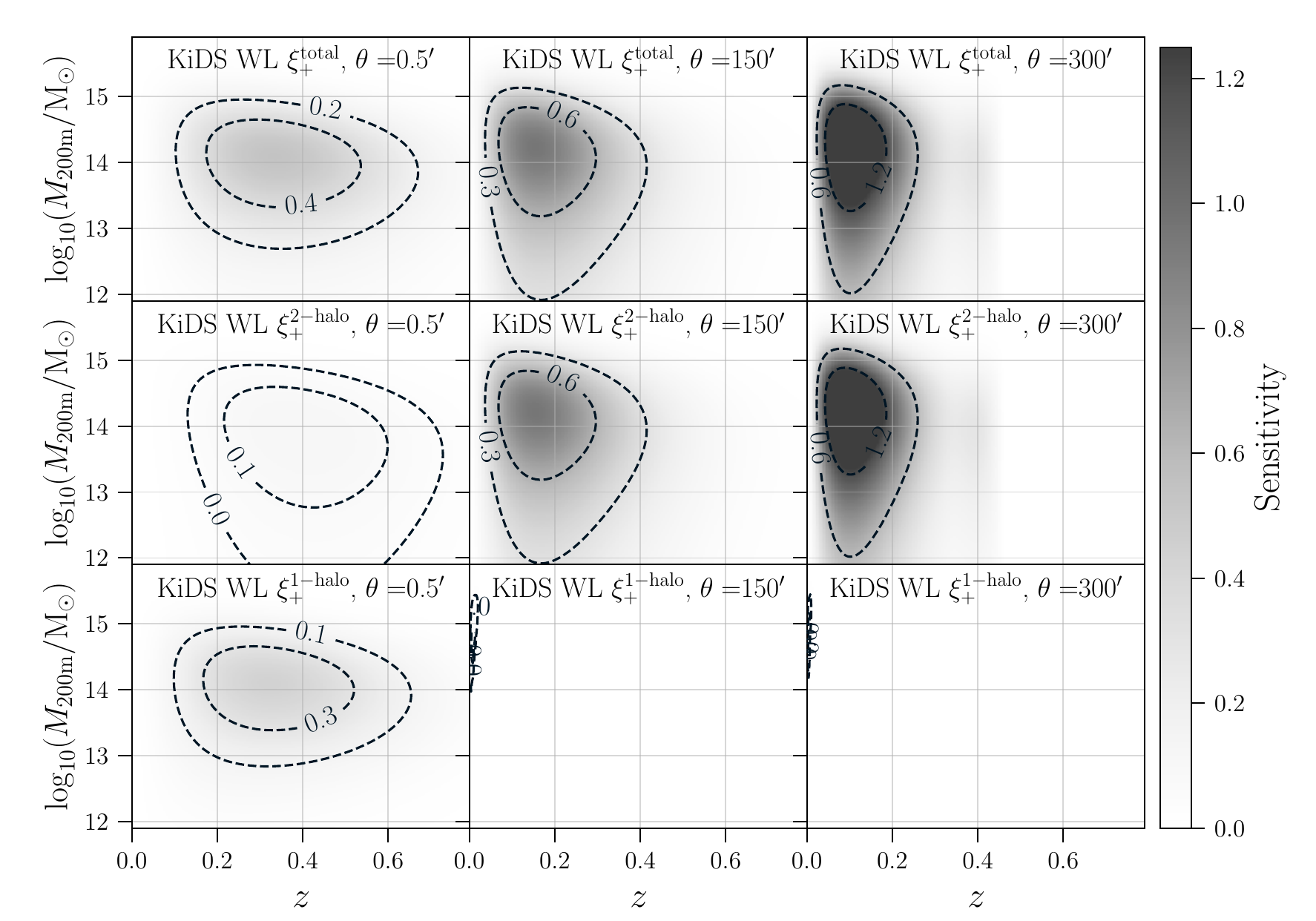}
    \includegraphics[width=0.49\textwidth]{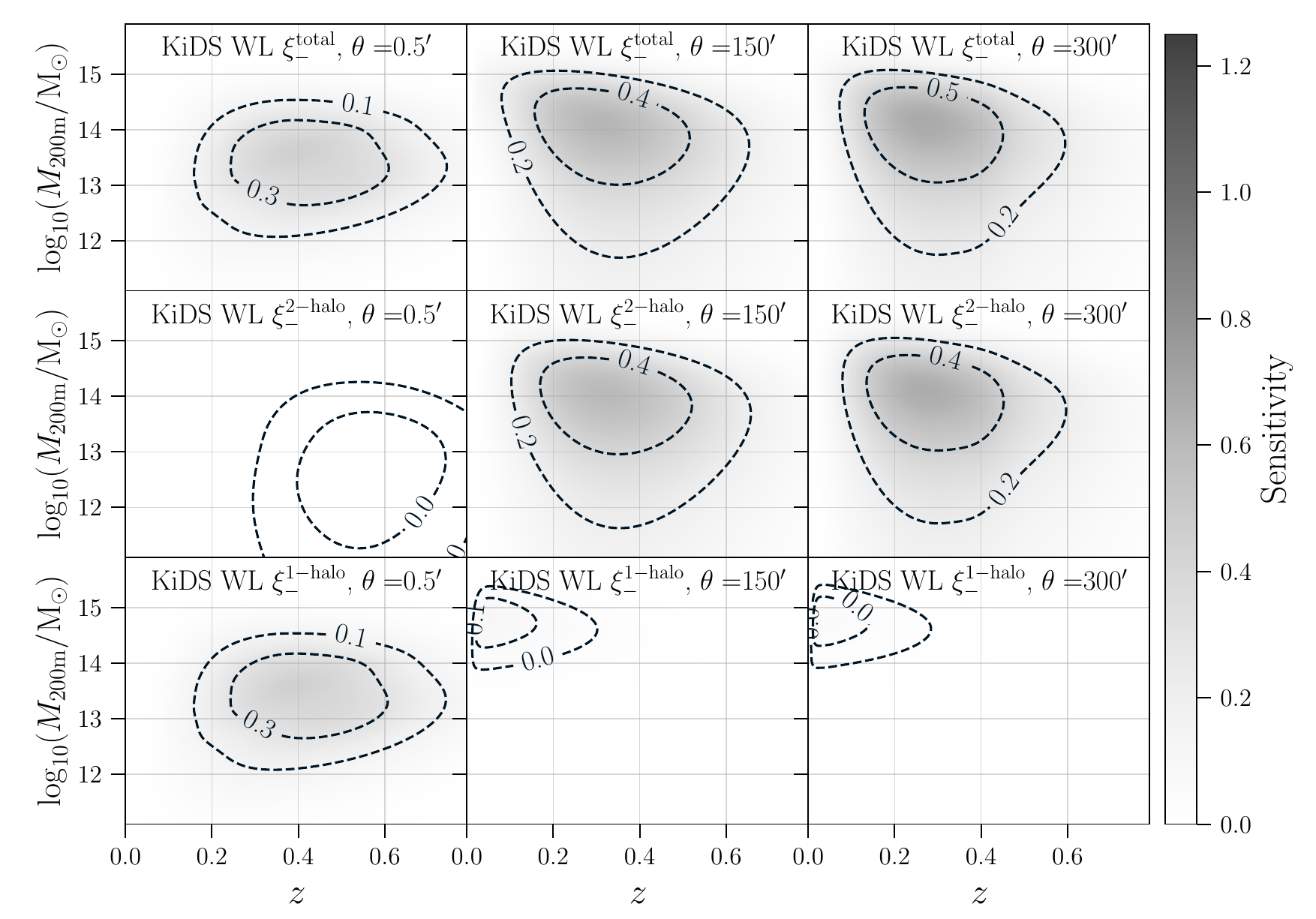}
    \caption{Sensitivity of cosmic shear $\xi_\pm$ for three representative scales (three columns from left to right in each sub-plot) corresponding to the smallest, middle, and largest angular scales at which the DESY3 (upper panels) and KiDS-1000 (lower panels) perform their measurements. In contrast to Fig.~\ref{fig:WL_sensitivity}, here we consider the full range of scales for which the measurements are available, without imposing any scale-cuts. The three rows of each sub-plot from top to bottom show the sensitivity to the total, two-halo and one-halo contributions respectively.}
    \label{fig:WL_sensitivity_appendix}
\end{figure*}
\section{Weak lensing sensitivity}
\label{app:WL}

DESY3 \cite{Secco_2022, Amon_2022} and KiDS-1000 \cite{Asgari_2021} measure the $\xi_{\pm}(\theta)$ signal on angular scales ranging from $\theta^{\rm DESY3}$ = $2.5$ to $200$ arcminutes and from $\theta^{\rm KiDS-1000} = 0.5$ to $300$ arcminutes, respectively. After imposing scale-cuts on these measurements to mitigate small-scale modeling uncertainties, the sensitivity on the lowest angular scales included by the two surveys for their cosmic shear cosmological analyses is shown in Fig.~\ref{fig:WL_sensitivity} of the main text. 

Fig.~\ref{fig:WL_sensitivity_appendix} shows the total, two-halo and one-halo sensitivity, in the top, middle, and bottom row of each sub-plot respectively, this time without considering any scale-cuts. In each row, we present results for three representative measurement scales corresponding to the smallest, middle and largest angular separations at which DESY3 (2.5, 127, 250 arcmin in their tomographic bin 4) and KiDS-1000 (0.5, 150, 300 arcmin in their tomographic bin 5) perform their $\xi_{\pm}$ measurements. The results for DESY3 are in the top panels ($\xi_+$ on the left and $\xi_-$ on the right), while those for KiDS-1000 are in the lower panels. 

The $\xi_{\pm}$ sensitivity trends (for both DESY3 or KiDS) show that on the smallest measured scales, the sensitivity is mainly dominated by the one-halo term. By contrast, for larger angular scales the two-halo contribution becomes significant, with the one-halo sensitivity becoming negligible. This further supports the motivation for imposing scale-cuts to mitigate uncertainties in the modeling of the one-halo term. The latter can arise either due to small-scale non-linear physics from baryonic feedback, or non-standard dark-matter dynamics that may impact the power spectrum predominantly in the one-halo regime. If one nevertheless chooses to consider small angular scales less than $\sim$ 5 arcmin (without scale-cuts or marginalizing over modeling uncertainties), the cosmic shear 1-halo term starts gaining sensitivity to lower halo masses, overlapping to some extent with DESI LRGs.

\bibliography{Feedback_MAH}
\end{document}